\documentclass[useAMS,usenatbib,usegraphicx]{mn2e}

\newcommand{\asec}{$^{\prime\prime}$}
\newcommand{\pas}{.\hskip-2pt$^{\prime\prime}$}
\def\IRAS{IRAS 20343+4129}
\def\ii{I20343}
\def\H{N$_{2}$H$^{+}$}

\def\AMM{NH$_3$}

\def\CO{\mbox{$^{12}$CO}}

\def\cdh{C$_2$H}
\def\nhdd{NH$_2$D}
\def\cthd{{\it c-}C$_3$H$_2$}
\def\HII{H{\sc ii}}

\def\UC{UC~H{\sc ii}}
\def\UCs{UC~H{\sc ii}'s}
\def\kms{\mbox{km~s$^{-1}$}}
\def\cmc{cm$^{-3}$}
\def\cmq{cm$^{-2}$}

\def\solm{\mbox{M$_\odot$}}
\def\soll{\mbox{L$_\odot$}}
\def\Vlsr{$V_{\rm LSR}$}

\def\Tk{\mbox{$T_{\rm k}$}}
\def\Tr{\mbox{$T_{\rm rot}$}}

\title[Dense gas in IRAS 20343+4129]{Dense gas in IRAS 20343+4129: an ultracompact \HII\ region caught in the act of creating a cavity}
\author[F. Fontani et al.]{F. Fontani$^{1}$\thanks{E-mail:
fontani@arcetri.astro.it}, Aina Palau$^{2}$, G. Busquet$^{3}$, A. Isella$^{4}$, R. Estalella$^{5}$, \'A. Sanchez-Monge$^{1}$,
\newauthor
P. Caselli$^{6}$ and Q. Zhang$^{7}$
\\
\\
$^{1}$INAF-Osservatorio Astrofisico di Arcetri, L.go E. Fermi 5, Firenze, I-50125, Italy\\
$^{2}$ Institut de Ci\`encies de l'Espai (CSIC-IEEC), Campus UAB-Facultat de Ci\`encies,Torre C5-parell 2, Bellaterra, E-08193, Catalunya, Spain\\
$^{3}$ INAF-Istituto di Fisica dello Spazio Interplanetario, Via Fosso del Cavaliere 100, Roma, I-00133, Italy\\
$^{4}$ Division of Physics, Mathematics and Astronomy, California Institute of Technology, MC 249-17, Pasadena, CA 91125, USA \\
$^{5}$ Departament de Astronomia i Meteorologia (IEEC-UB), Institut de Ci\`encies del Cosmos, Universitat de Barcelona, Marti Franqu\`es 1,\\ 
E-08028 Barcelona, Spain \\
$^{6}$ School of Physics and Astronomy, University of Leeds, Leeds, LS2 9JT , UK\\
$^{7}$ Harvard-Smithsonian Center for Astrophysics, 60 Garden Street MS78, Cambridge, MA 02138, USA \\
}
\begin{document}

\date{Accepted date. Received date; in original form date}

\pagerange{\pageref{firstpage}--\pageref{lastpage}} \pubyear{2011}

\maketitle

\label{firstpage}

\begin{abstract}
The intermediate- to high-mass star-forming region \IRAS\ is an excellent laboratory 
to study the influence of high- and intermediate-mass young stellar
objects on nearby starless 
dense cores, and investigate for possible implications in the clustered star 
formation process.
We present 3~mm observations of continuum and rotational transitions of 
several molecular species 
(\cdh , \cthd , \H , \nhdd) obtained with the Combined Array for Research 
in Millimetre-wave Astronomy, as well as 1.3~cm continuum and \AMM\
observations carried out with the Very Large Array, to reveal the 
properties of the dense gas. We confirm undoubtedly previous claims 
of an expanding cavity created by an ultracompact \HII\ region
associated with a young B2 zero-age main sequence (ZAMS) star. The dense 
gas surrounding the cavity
is distributed in a filament that seems squeezed in between the cavity 
and a collimated outflow associated with an intermediate-mass
protostar.
We have identified 5 millimeter continuum condensations in the filament.
All of them show column densities consistent with potentially being the
birthplace of intermediate- to high-mass objects.
These cores appear different from those 
observed in low-mass clustered environments in 
sereval observational aspects (kinematics, temperature, 
chemical gradients), indicating a strong influence of the most 
massive and evolved members of the protocluster.
We suggest a possible scenario in which the B2 ZAMS star
driving the cavity has compressed the surrounding gas, perturbed
its properties and induced the star formation in its immediate surroundings. 

\end{abstract}


\section{Introduction}
\label{intro}

Most of the stars of all masses in the Galaxy form in rich clusters.
Despite this, the details of the clustered star formation process are still poorly understood. 
%
Studies of {\it low-mass protoclusters} have started to unveil similarities and 
differences between isolated and clustered dense cores (e.g.~
Andr\'e et al.~\citeyear{andre}, Foster et al.~\citeyear{foster}, 
Friesen et al.~\citeyear{friesen}).
%
%
Globally these studies suggest that cluster environment has a relatively smaller 
influence on the properties of the cores (temperature, mass, velocity dispersion,
chemical abundances of early phase molecules) than is typically assumed (Foster et al.~2009).
However, the conclusions described above do not include observations of
{\it high-mass star forming regions}.
Because the phenomena associated with massive star formation have a stronger
impact on the environment (massive outflows, UV radiation, expanding \HII\ regions), 
it is plausible that these energetic phenomena have major effects on 
the surrounding dense material.
The study of such interaction is especially important to 
quantify the effect of protostellar feedback on the environment and
test recent models of high-mass star formation including outflows
and radiation from the newly born stars (e.g.~Krumholz et al.~2011,
Hennebelle et al.~\citeyear{hennebelle}).

\begin{figure*}
\centerline{\includegraphics[angle=0,width=16cm]{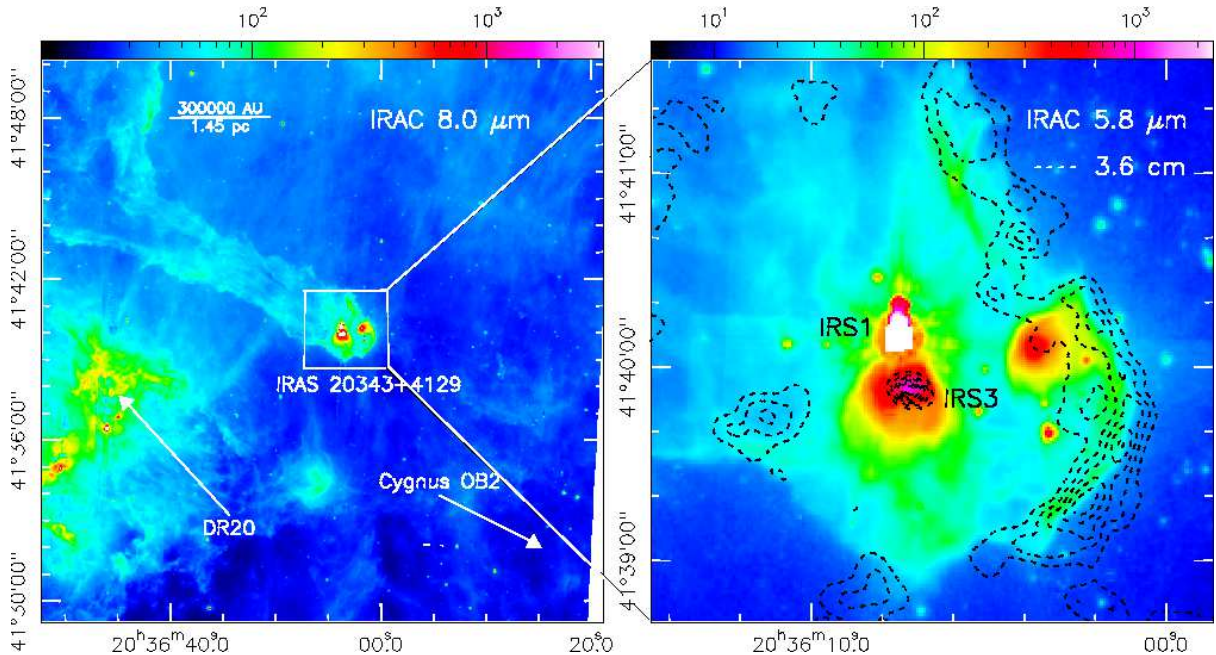}}
\caption{Large-scale view of the surroundings of \IRAS\ as seen by Spitzer at 8.0~$\mu$m.
The Cygnus OB2 association is located to the south-west as shown by the arrow.
The enlargement panel to the right shows the region of our interest as it appears 
in the Spitzer image at $5.8$~$\mu$m, where we highlight 
the two infrared sources IRS~1 (saturated in the IRAC image) and IRS~3. 
In both panels, the colour scale units are MJy sr$^{-1}$.
The dashed contours correspond
to the 3.6~cm continuum emission detected by Carral et al.~(\citeyear{carral})
with the VLA (D-configuration).
}
\label{fig_largefov}
\end{figure*}

Interferometric observations of dense gas and dust tracers (\H , mm continuum, \AMM )
have revealed the presence of pre-stellar core candidates 
surrounding ultracompact \HII\ regions (\UCs ) and other massive 
young stellar objects (YSOs) that do show evidences of such an interaction.
For example, Fontani et al.~(2009) found that 
in the protocluster associated with IRAS 05345+3157 the kinematics of two pre-stellar 
core candidates is influenced by the passage of a massive outflow.  UV radiation 
and powerful outflows affect the chemistry of starless cores in
IRAS 20293+3952 (Palau et al.~2007a). On the other hand, a crucial chemical 
process in pre--stellar cores, i.e. the deuteration of species like \H\ and \AMM ,
seems to remain as high as in pre-stellar cores isolated and associated 
with low-mass star forming regions (Fontani et al.~\citeyear{fonta08}, 
Busquet et al.~2010, Pillai et al.~\citeyear{pillai11}). Therefore, to date it is not
clear if and how the presence of massive objects affects
the properties and evolution of the other (pre-)protocluster members.

The protocluster associated with \IRAS\ (hereafter \ii) represents an excellent laboratory 
to study this issue.
The IRAS source is located to the northeastern side of the Cygnus OB2 association, 
at 1.4 kpc of distance from the Sun (Sridharan et al.~2002, Rygl et al.~\citeyear{rygl}), 
and two bright nebulous stars, 
IRS 1 (north) and IRS 3 (south), are found inside the IRAS error ellipse when
observed at high angular resolution (Kumar et al.~\citeyear{kumar}). 
The bright infrared stars are embedded in a cometary-like cloud whose head, 
facing the Cygnus OB2 association, is bright at centimeter wavelengths 
and whose tail, bright in the mid-infrared, is extending for about $10'$ ($\sim4$~pc) 
towards the north-east (Fig.~\ref{fig_largefov}). This kind of clouds are also
known as bright rimmed clouds.

Thanks to interferometric observations of \CO\ and 1.3~mm continuum, 
Palau et al.~(\citeyear{palau07b}) concluded that IRS 1 is an intermediate-mass Class I YSO 
driving a molecular outflow in the east-west direction, while IRS 3 is likely a 
more evolved intermediate/high-mass star. 
This is further confirmed through mid-infrared photometric and spectroscopic observations, 
from which Campbell et al.~(\citeyear{campbell}) also estimated a bolometric luminosity 
of the order of 1000~\soll\ for both IRS 1 and IRS 3. Furthermore, IRS 3 is at the centre of 
an UC \HII -region detected through VLA centimeter continuum
emission (Carral et al.~\citeyear{carral}), and of a fan-shaped emission 
in the 2.12~$\mu$m rovibrational line of molecular hydrogen (Kumar et al.~\citeyear{kumar}). 
East and west of this fan-shaped feature, Palau et al.~(\citeyear{palau07b}) detected  
molecular gas and dust resolved into several millimeter continuum compact sources. 
Palau et al.~(\citeyear{palau07b}) interpreted these starless condensations 
as being accumulated on the walls of the expanding shock front, but could
not derive firm conclusions on their origin and nature. 


This work aims at better understading the nature of the dense cores in \ii , and its relation
with the neighbouring more evolved objects. To achieve the goal we performed 
observations of molecular species obtained at high angular resolution with the 
Combined Array for Research in Millimeter Astronomy (CARMA) at 
3~mm and the Very Large Array (VLA) at 1.3~cm. 
All selected molecular transitions are commonly used to characterise
dense gas: (i) \AMM\ and \H\ are excellent tracers of dense and 
cold cores because either species do not suffer from
depletion up to $\sim 10^5$ \cmc , and \AMM\ is extensively
used as thermometer in both low- and high-mass star forming
regions; (ii) \nhdd\ provides an estimate of the degree of 
deuteration (with \AMM ). This combination of diagnostic lines was 
successfully used by Busquet et al.~(2010) to identifying pre-protostellar cores in 
the proto-cluster associated with IRAS 20293+3952; 
(iii) C$_2$H and \cthd\ are both high-density PDR tracers useful to shed light on the interaction
among the cold gas and the UV radiation field coming from IRS 1 and IRS 3.
\cdh\ is also a tracer of cold gas (e.g. Beuther et al.~2008, Padovani et al.~\citeyear{padovani}).
In this paper we concentrate on the gas morphology, temperature and 
kinematics of the region adjacent to IRS 1 and IRS 3, 
and confirm the hypothesis proposed by Palau et al.~(\citeyear{palau07b}) 
that IRS 3 is opening a cavity in the surrounding dense gas and 
starless material is being accumulated on the cavity walls.
In Sect.~\ref{obs} we describe the observations. The observational
results are presented in Sect.~\ref{res}, and discussed in 
Sect.~\ref{discu}. In Sect.~\ref{conc} we summarise the main findings
of the work and give a general conclusion.

\section{Observations and data reduction}
\label{obs}

\subsection{CARMA}
\label{carma}

3~mm CARMA observations of \ii\ were obtained on 29 Mar 2010 in C- and
01 May 2010 in D-configuration under good weather conditions for 
observations at 3 mm,  characterized by about 5 mm of precipitable water 
and atmospheric noise rms of about 300 $\mu$m as measured on a baseline 
of 100 m at the frequency of 225 GHz.  The phase centre was the same as in
Palau et al.~(\citeyear{palau07b}), namely: 
R.A. (J2000) = $20^{\mathrm h}36^{\mathrm m}07\fs3$ and 
Dec. (J2000) = $41^{\circ }39^{\prime }57$\pas 2.
The local standard of rest velocity of the cloud is assumed to
be 11.5 \kms , as determined
from single-dish ammonia observations (Sridharan et al.~2002).
The primary beam of the 10~m and 6~m dishes at about 85~GHz 
is $\sim 73$ \arcsec\ and 121 \arcsec , respectively.
The single-side-band system temperature 
during the observations was below 150 K. During C-configuration observations, the 
correlator provided 4 bands which were configured to observe 
the continuum, the \cdh , the \nhdd\ and the \H\ lines simultaneously. 
D-array observations were 
obtained with the new CARMA correlator, which provides more bands. Two 500 MHz
bands were used to observe the continuum and 5 bands set up 
to observe the \cdh , \H , \nhdd , CCS and \cthd\ line emission.
The pass-band was calibrated by observing 1733-130; flux calibration was
set by observing MWC349. The estimated uncertainty of the absolute flux 
calibration is 10\% , and it is determined from periodic observations
of MWC349. Atmospheric and instrumental effects were corrected by 
observing the nearby quasar 2007+404 every 15 minutes.

The tracers observed and the main observational 
parameters (frequency, synthesised beam, linear resolution, spectral resolution, 
1$\sigma$ rms channel noise, largest detectable angular scale) are reported in Table~\ref{obs_par}. 
The CCS line is the only undetected transition and will be not discussed
in the following.
The continuum was derived by averaging the 500 MHz bands.
Visibility data were edited and calibrated with the MIRIAD package. 
A minor flagging of the data was performed using the UVFLAG task, 
mainly to remove the intervals characterized by
the bad atmospheric phase coherence. The channel
spacing and the corresponding 1 $\sigma$ rms noise are shown in Table~\ref{obs_par}. 
Merging the visibilities obtained in C and D configuration, imaging, 
deconvolution, and analysis of channel maps and continuum
were performed using the standard tasks of the 
GILDAS\footnote{the GILDAS software is developed 
at the IRAM and the Observatoire de Grenoble, and is available at 
http://www.iram.fr/IRAMFR/GILDAS} package (e.g. UVMERGE, UVMAP, CLEAN).
Images were created applying natural weighting to the visibilites.

\begin{table*}
\centering
\begin{minipage}{160mm}
\caption[] {Observed tracers and basic parameters for the CARMA and VLA observations. 
The synthesised beam and 1 $\sigma$ rms for CARMA observations are based on the 
combined configurations C+D (unless when differently specified)}
\label{obs_par}
\begin{tabular}{lcccccc}
\hline \hline
Instrument/Tracer \footnote{For the molecular transitions, in the text we will use the following abbreviations: \nhdd\ $(1_{1,1}-1_{0,1})$ = \nhdd\ (1--1); 
\cdh\ $(1_{3/2,2}-0_{1/2,1})$ = \cdh\ (1--0); \cthd\ $(2_{1,2}-1_{0,1})$ = \cthd\ (2--1); \H\ $(1_2-0_1$) = \H\ (1--0);}
&Frequency \footnote{rest frequency of the transition listed in Col.~1;}
&Synth.\ beam
&Linear
& $\Delta v$
&1$\sigma$ rms
&LAS\footnote{largest angular scale (at half power) detectable by the interferometer, estimated from the minimum baseline of the array configuration, and following the appendix in Palau et al. (\citeyear{palau10}). For the lines observed in C and D configuration, this refers to the merged uv coverage;}
\\
&(GHz)
&(\asec$\times$\asec )
&resolution (pc)
& (\kms)
&(Jy~beam$^{-1}$)
&($''$)
\\
\hline 
\emph{CARMA}\\
\hline
3~mm continuum
	&86.4197	& $4.80\times4.36$	&$\sim$0.03	&  --	&0.0004	& 33	\\
\nhdd\ $J_{K_{a},K_{c}} = 1_{1,1}-1_{0,1}$ ($F=1-1$)
	&85.9263	&$3.32\times2.68$	&$\sim$0.02	&0.07 & 0.045\footnote{sensitivity in the merged C+D channel maps smoothed to a spectral resolution of 0.1 \kms ;}	&33	\\
\cdh\ $N_{J,F} = 1_{3/2,2}-0_{1/2,1}$
	&87.3169	&$3.35\times2.72$	&$\sim$0.02	& 0.41  & 0.02	&33	\\
{\it ortho-}\cthd\ $J_{K_{a},K_{c}} = 2_{1,2}-1_{0,1}$\footnote{observed in D configuration only; }
	&85.3389	&$5.81\times4.54$	&$\sim$0.034	&0.07	&0.07	& 33\\
CCS $N_{J} = 7_{6}-6_{5}$
	&86.1814	&$5.67\times4.84$	&$\sim$0.035	&0.07	&0.04	&33	\\
\H\ $J_{F_1} = 1_2-0_1$ ($F=3-2$)\footnote{observed in C configuration only.}
	&93.1737	&$2.04\times1.94$	&$\sim$0.013	&0.065	&0.07	& 16 \\
\hline 
\emph{VLA}\\
\hline
7~mm continuum
	&43.3399		&$4.67\times4.57$	&$\sim$0.03	&--	&0.0004	&18\\
1.3~cm continuum
	&22.4601		&$3.02\times2.98$	&$\sim$0.02	&--	&0.000057&34\\
NH$_3$\,(1,1)
	&23.6945		&$4.20\times3.19$	&$\sim$0.025	&0.62&0.0035	&$34$\\
NH$_3$\,(2,2)
	&23.7226		&$4.25\times3.19$	&$\sim$0.025	&0.62&0.0035	&$34$\\
\hline
\end{tabular}
\end{minipage}
\end{table*}


\subsection{VLA}
\label{vla}

I20343 was observed with the Very Large Array (VLA\footnote{The Very 
Large Array (VLA) is operated by the National Radio Astronomy Observatory (NRAO), a 
facility of the National Science Foundation operated under cooperative agreement by
Associated Universities, Inc.} at 1.3 and 0.7~cm on 2007 Mar 26 using 
the array in the D configuration\footnote{The VLA continuum observations presented 
in this work correspond to project AP525. Another project, AP533, was undergone 
at 3.6, 2 and 0.7~cm, but due to technical problems the data of AP533 project were
lost.}. The phase center of the observations was R.A.\ (J2000) =
20$^\mathrm{h}$36$^\mathrm{m}$07$\farcs$51; Dec.\ (J2000) =
41$^{\circ}$40$^\prime$00\pas9. The data reduction followed the VLA standard
guidelines for calibration of high frequency data, using the NRAO 
package AIPS.
The absolute flux scale was set by observing the quasar 1331+305 
(3C286), for
which we adopted a flux of 2.52~Jy at 1.3~cm, and 1.45~Jy at 0.7~cm. The 
quasar
2015+371, with a bootstrapped flux of $1.39\pm0.02$~Jy at 1.3~cm and
$2.1\pm0.2$~Jy at 0.7~cm, was observed regularly to calibrate the gains and
phases. Final images were produced with the robust parameter of Briggs 
(1995)
set to 5, corresponding to natural weighting. At 7~mm we applied a taper at
60~k$\lambda$ with the aim of recovering faint extended emission, but no
emission was detected at this wavelength.

The VLA was also used to map the ($J$,$K$)=(1,1) and (2,2) inversion transitions of
the ammonia molecule on 2001 July 23, with the array in the C configuration.
The phase center was set to R.A.(J2000) $=20^{\mathrm h}36^{\mathrm m}08\fs013$;
Dec.(J2000)$=+41\degr39\arcmin56\farcs93$. 
The FWHM of the primary beam at the observed frequency was
$\sim$110$''$, and the range of projected baselines was 2.59 to 267.20~k$\lambda$. 
The absolute flux calibration was performed by using 3C286, adopting a 
flux density at 1.3 cm of 2.41 Jy.
The phase calibrator was QSO 2013+370, with a 1.3 cm bootstrapped flux 
density of 2.34$\pm$0.04 Jy, and 3C273 was used as the
bandpass calibrator. The \AMM(1,1) and \AMM(2,2) lines were observed simultaneously in the 4 IF
correlator mode of the VLA (with 2 polarizations for each line), providing 63
channels with a spectral resolution of 0.62~\kms \ across a bandwidth of
3.13~MHz, plus an additional continuum channel containing the central 75\% of
the total bandwidth. The bandwidth was centered at the systemic velocity
\Vlsr=11.5~\kms\ (Sridharan et al. 2002) for the \AMM(1,1) line, and
at \Vlsr=6.5~\kms \ for the \AMM(2,2) line (to cover the main and
one of the satellite components). Data were calibrated using
standard procedures of AIPS, and imaging was performed by applying a taper 
in the $uv$-data of 50~k$\lambda$ and using natural weighting to recover the faint and extended emission. 
The synthesised beams and rms noises for all the VLA observations are listed in Table~\ref{obs_par}.

\section{Results}
\label{res}

\subsection{Continuum emission maps}
\label{continuum}

\begin{figure}
\centerline{\includegraphics[angle=0,width=8cm]{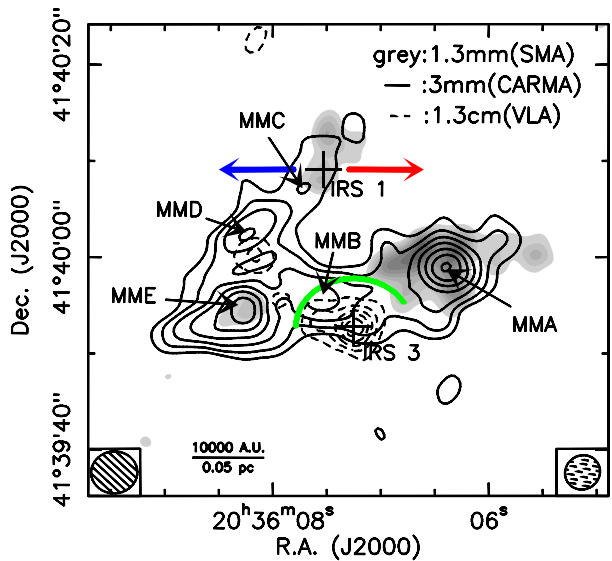}}
\caption{Continuum emission obtained with CARMA at 3~mm (solid contours)
and with the VLA at 1.3~cm (dashed contours)
towards \ii . The solid contours start from the 3$\sigma$ rms level (1.2$\times 10^{-3}$ Jy beam$^{-1}$)
and are in steps of 1$\sigma$ rms. The dashed first contour and step correspond
to the 3$\sigma$ rms level (1.5$\times 10^{-5}$ Jy beam$^{-1}$)
and are in steps of 1$\sigma$ rms
The grey scale indicates the 1.3~mm continuum
emission observed by Palau et al.~(\citeyear{palau07b}) with the SMA (levels = 10$\%$ of the
maximum, first level = 20$\%$ of the maximum). The crosses indicate the two
infrared sources idenfitifed by Kumar et al.~(\citeyear{kumar}).
Red and blue arrows highlight the direction of the blue- and red-shifted emission
detected in \CO\ (2-1) by Palau et al.~\citeyear{palau07b}. The green arc
roughly shows the fan-shaped feature detected in H$_2$ $\mu$m associated with IRS 3
(Kumar et al.~\citeyear{kumar}), interpreted as an expanding cavity by
Palau et al.~\citeyear{palau07b}. The ellipse at bottom-left corresponds to
the CARMA synthesised beam (4.8\asec $\times$ 4.4\asec , at P.A. $92^{o}$).
The dashed ellipse at bottom-right is the VLA synthesised beam 
(3.71\asec $\times$ 3.50\asec , at P.A. $-56.2^{o}$).
}
\label{fig_cont}
\end{figure}

In Fig.~\ref{fig_cont} the 3~mm (CARMA, C+D configuration, solid contours) 
and 1.3~cm (VLA, D configuration, dashed contours) continuum emission maps are shown. 
As reference, we overplot the 1.3~mm 
continuum emission observed by Palau et al.~(\citeyear{palau07b}, obtained
with about a factor 1.5 better angular resolution) as well as the direction of the lobes of 
the \CO\ outflow originated by IRS 1 and the (rough) edge of the fan-shaped
H$_2$ emission associated with IRS 3 (Kumar et al.~\citeyear{kumar}).

The 3~mm continuum emission is resolved into 5 main condensations,
which we call MMA, MMB, MMC, MMD and MME in order of increasing R.A.
The brightest are MME and MMA, east and west of IRS~3, respectively. 
From Fig.~\ref{fig_cont} we note that MME roughly coincides with a 
1.3~mm continuum peak (Palau et al.~\citeyear{palau07b}), 
while the westernmost one, MMA, encompasses 3 peaks seen at
1.3~mm with SMA. 
The faintest core, MMC, is detected towards IRS~1. 
Two more 3~mm condensations, MMD and MMB, are not detected at 1.3~mm.
Specifically, MMB is clearly detected close to IRS3, with a shift of only 4\arcsec\ 
to the northeast.

The most evident differences among the 3~ and 1.3~mm continuum maps are that
at 3~mm core MME (MM7 in Palau et al.~\citeyear{palau07b}) is more extended,
and the eastern and western cores are connected by a filamentary emission 
passing through IRS~3 which is undetected at 1.3~mm. Both differences
are probably just the consequence
of CARMA being sensitive to larger structures than the SMA.  
We estimated that the SMA in the observations of Palau et al.~(\citeyear{palau07b}) 
was sensitive to structures with FWHM $<9''$ (using the minimum baseline of the 
observations and following Palau et al.~\citeyear{palau10}), while CARMA using C+D 
configurations was sensitive to structures $<33''$ (Table~\ref{obs_par}), 
allowing CARMA to recover more extended emission. 


As shown in Fig.~\ref{fig_cont}, the continuum emission at 1.3~cm is dominated by one strong 
and compact source with its emission peak (20$^\mathrm{h}$36$^\mathrm{m}$07.3$^\mathrm{s}$,
+41$\degr$39'52") coincident with IRS~3, and matching well the
fan-shaped H$_2$ emission detected by Kumar et al. \citeyear{kumar}. 
A Gaussian fit to this compact centimeter source yields a peak intensity of 
0.8~mJy/beam, a flux density of 1.3~mJy, and a deconvolved size of $4.3''\times1.0''$, 
with PA=78\degr, corresponding to a size of 6000 AU in the east-west direction 
(and an unresolved size in the north-south direction). 
In addition to the compact 1.3~cm source associated with IRS~3, there is a secondary 
peak at around 6$\sigma$ which falls $2''$ to the south of MMD, and faint emission 
joining this secondary peak and the centimeter source in IRS~3, 
suggesting that the two peaks of centimeter emission could be linked.

\subsection{Distribution of the Integrated intensity of molecular line emission}
\label{distribution}

\subsubsection{Molecular tracers observed with CARMA}
\label{res_linescarma}

\begin{figure*}
\centerline{\includegraphics[angle=0,width=15cm]{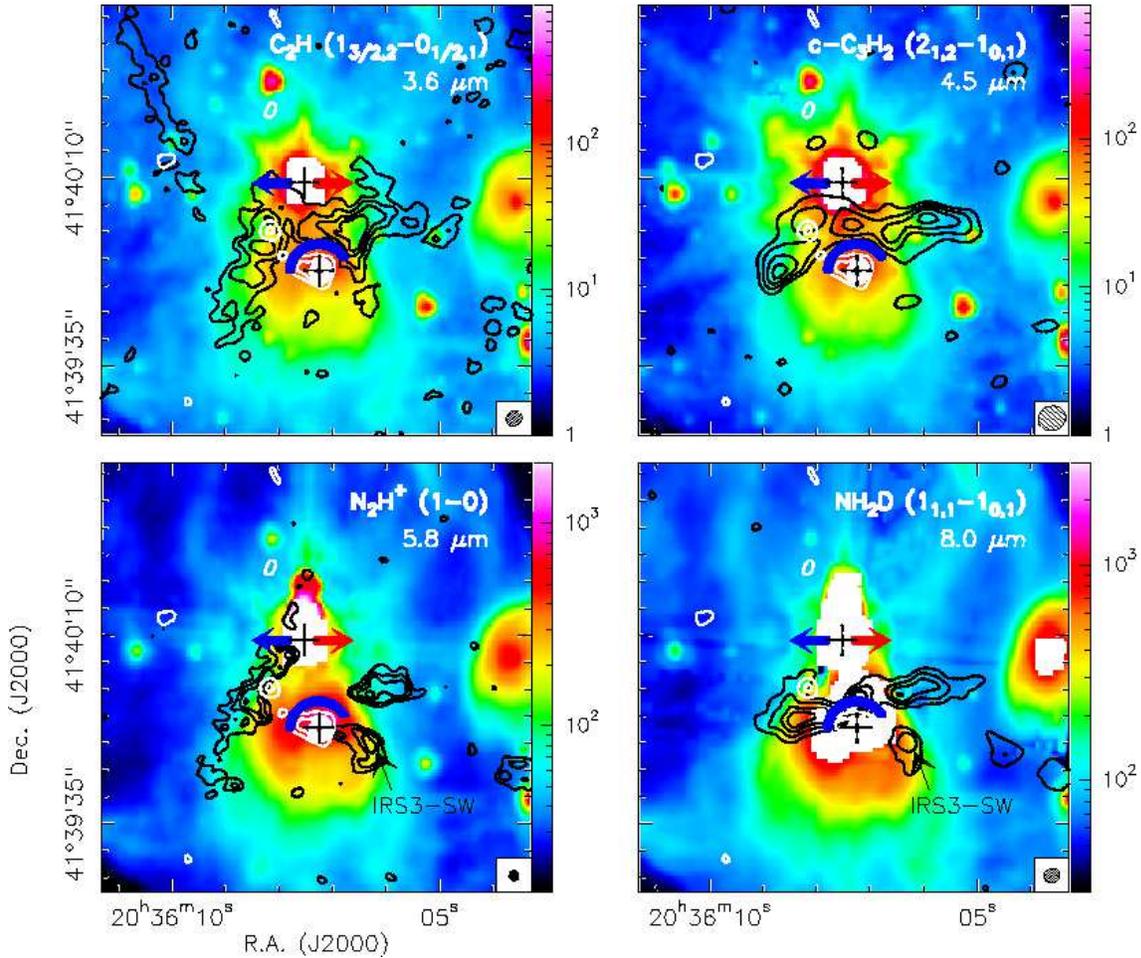}}
\caption{Integrated maps of the molecular lines detected with CARMA 
(Table~\ref{obs_par}) towards \ii\ and superimposed 
on the images obtained in the Spitzer-IRAC
bands (3.6 $\mu$m, 4.5 $\mu$m, 5.8 $\mu$m, 8 $\mu$m,
colour scale in MJy sr$^{-1}$ units). 
In each panel, black contours depict the velocity-averaged
emission of the transition labelled at the top-right corner (the corresponding 
synthesised beam is shown in the bottom-right corner). 
For all lines, the emission has been averaged over all the velocity 
channels with signal, except for \H\ (bottom-left panel) for which 
the integrated emission was averaged over the main group of hyperfine 
components only. 
For C$_2$H and NH$_2$D, first contour and step correspond to the 20$\%$ 
level of the maximum (corresponding roughly to the 3$\sigma$ rms level
of the averaged map), while for \cthd\ and \H\ contours start from the 30$\%$ 
level of the maximum and are in steps of 20$\%$.
The white contours represent the VLA 1.3~cm (K-band) continuum image
(same contours as in Fig.~\ref{fig_cont}).
The position of IRS 1 and IRS 3, the associated outflow and fan-shaped
H$_2$ emission are shown as in Fig.~\ref{fig_cont}.}
\label{fig_CARMA_tot}
\end{figure*}

The maps of the integrated intensity of the lines observed and detected with 
CARMA (see Table~\ref{obs_par}) are shown in Fig.~\ref{fig_CARMA_tot}.
The emission map of each tracer has been superimposed on the 
images obtained from the Spitzer Space Telescope in the four mid-IR
IRAC bands (centred at 3.6, 4.5, 5.8 and 8 $\mu$m, respectively). 
The location of the near-infrared sources IRS 1 and IRS 3 are also indicated, 
as well as the direction of the outflow centred on IRS 1 and the edge of
the H$_2$ emission associated with IRS 3, as in Fig.~\ref{fig_cont}.
We also superimpose the 1.3~cm continuum emission detected in
this work (see Fig.~\ref{fig_cont}), which marks clearly the \HII\ region 
associated with IRS 3.

The molecular gas seems to be squeezed in between the 
two dominant mid-IR sources IRS 1 and IRS 3 in all molecules except
\H . Other compact mid-IR sources are 
present in the region but do not seem to be associated
with any clear molecular counterpart. The diffuse IR nebulosity, especially
evident in the 5.8 and 8 $\mu$m bands, is
probably emission from small dust grains distributed around \ii\
becoming brighter at longer wavelengths.
Some of the diffuse emission detected at 8 $\mu$m may
also be PAHs emission (e.g.~Peeters et al.~\citeyear{peeters}).

The morphology of the integrated intensity of \cthd\ (2-1) and \cdh\ (1-0) delineates 
clearly a cavity around IRS~3 (top panels of Fig.~\ref{fig_CARMA_tot}), 
providing a strong support to the hypothesis proposed 
by Palau et al.~(\citeyear{palau07b}), namely that IRS3 is most likely
a more evolved intermediate-mass star creating a cavity. The \cdh\ emission is more extended
than that of \cthd , perhaps due to the smaller sensitivity that
we have in the \cthd\ channel maps (see Table~\ref{obs_par}).
Specifically, a narrow filament extended $\sim 30$\arcsec\ 
is clearly detected north-east of the field center
(Fig.~\ref{fig_CARMA_tot}, top-left panel), inclined roughly as the
tail of the mid-IR cometary shape (see Fig.~1), suggesting that the two 
features can have the same origin.

The bottom panels in Fig.~\ref{fig_CARMA_tot} show the integrated 
emission of the two Nitrogen-bearing species \H\ and \nhdd . 
The emission in \H\ consists mainly of one cloud to the east of IRS~3 
elongated in the southeast-northwest direction, and extending up to IRS~1, 
and two smaller clouds, one immediately to the south-west of IRS~3 with
no continuum counterpart (called IRS3-SW), 
and the other associated with MMA. In addition, there is one clump about $1$\arcmin\ to the west of 
IRS~3, almost at the border of the bright rim, which falls on a region with no infrared 
emission associated. On the other hand, the emission of \nhdd\ consists mainly 
of one filament elongated in the east-west direction, passing through IRS~3, 
and with some emission at IRS3-SW. 
The \H\ clump $1'$ to the west of IRS~3 is detected also in \nhdd ,
but looks more extended in \nhdd . However,
this can be just an effect of the different angular resolution 
and different filtering of extended emission, as \nhdd\  was observed with 
C+D configuration while \H\ was observed in C configuration only (see Table~\ref{obs_par}).

\begin{figure}
\centerline{\includegraphics[angle=0,width=8cm]{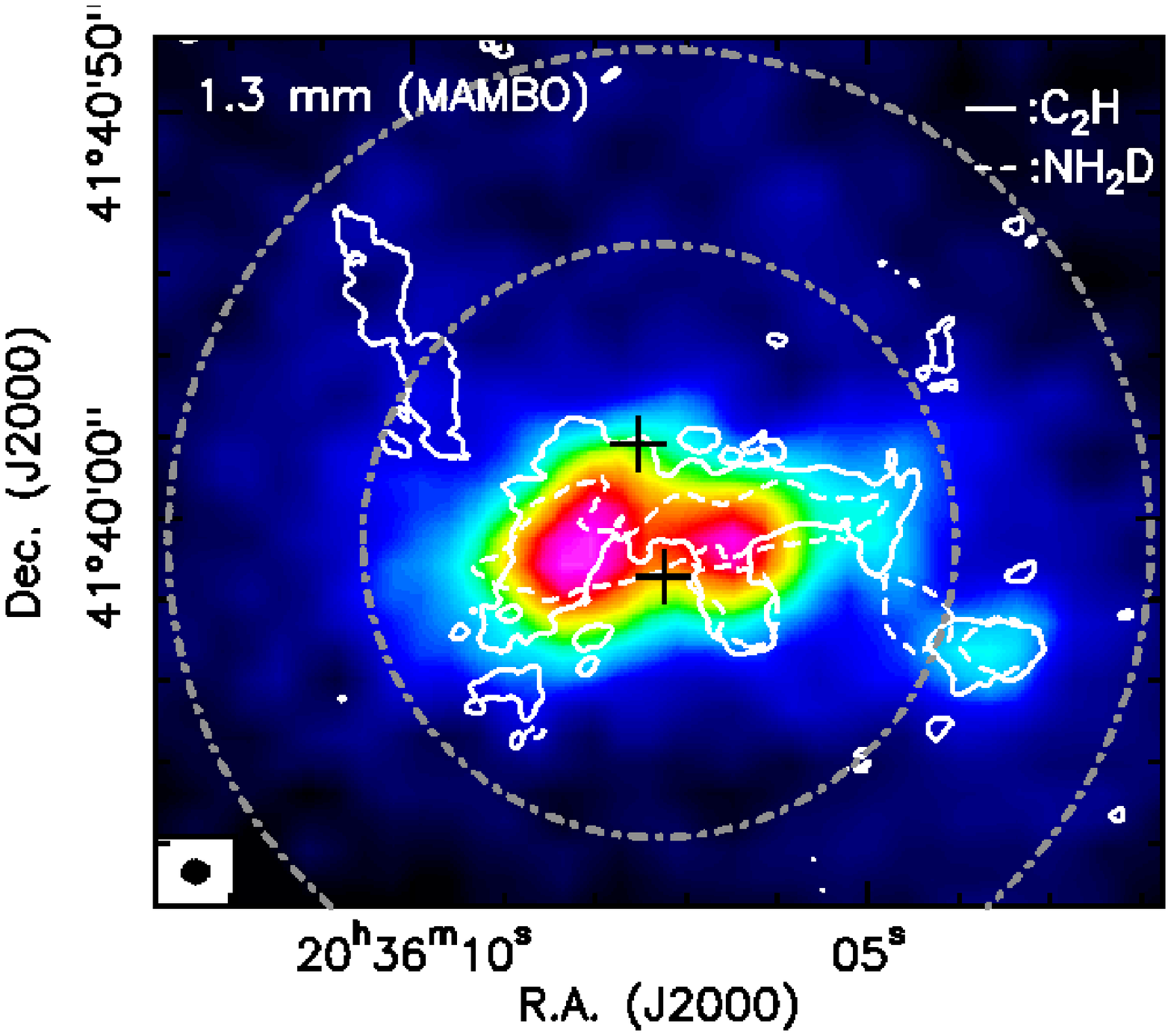}}
\caption{Integrated emission of molecular tracers (white contours) superimposed
on the  image of \ii\ in MAMBO at 1.2 mm (Beuther et al. 2002).
The solid white contour depicts the 15$\%$ level of the zero-order moment
map of \cdh\ (1--0). The dashed white contour corresponds to the 15$\%$ 
level of the zero-order moment map for the \nhdd\ (1--1) line. 
The ellipse in the bottom-left corner shows the CARMA synthesised 
beam for the \cdh\ map and it is almost coincident to that of the \nhdd\ map. 
The grey dotted-dashed circles represent the primary beams of the 
CARMA 10 and 6~m dishes (73 and 121 \arcsec ), respectively.}
\label{fig_mol_all}
\end{figure}

If we put together the two mostly extended molecular tracers,
namely \cdh\ and \nhdd , we can notice a sort of 'snake-like' filament
of molecular gas (Fig.~\ref{fig_mol_all}) extending from the south-western side
of \ii , clearly detected in \nhdd , up to the north-eastern corner, in which
a long and narrow filament is detected in \cdh . The bulk of the emission is 
in between IRS 1 and IRS 3. This 'snake-like' filament 
matches very well the 1.2~mm continuum emission detected with MAMBO
(Beuther et al.~2002, Fig.~\ref{fig_mol_all}), and its SW-NE inclination follows 
roughly the 'head-tail' orientation of the mid-IR diffuse cometary 
emission (Fig.~\ref{fig_largefov}), 
suggesting a possible common 
origin. However, the highest sensitivity region of the CARMA maps, 
i.e. the field of view of the 10~m dishes ($\sim 73$\arcsec ), includes only the central region 
of the filament (see Fig.~\ref{fig_mol_all}). For this reason, in this work we focus on the 
centre of \ii , where the interaction between the two brightest IR sources 
and the surrounding molecular gas seems predominant, and plan a 
large interferometric mosaic which will allow us to unveil the overall 
distribution of the molecular gas and its relation with all the IR sources.

\subsubsection{Ammonia (1,1) and (2,2) inversion transitions}
\label{res_linesvla}

\begin{figure}
\centerline{\includegraphics[angle=0,width=8cm]{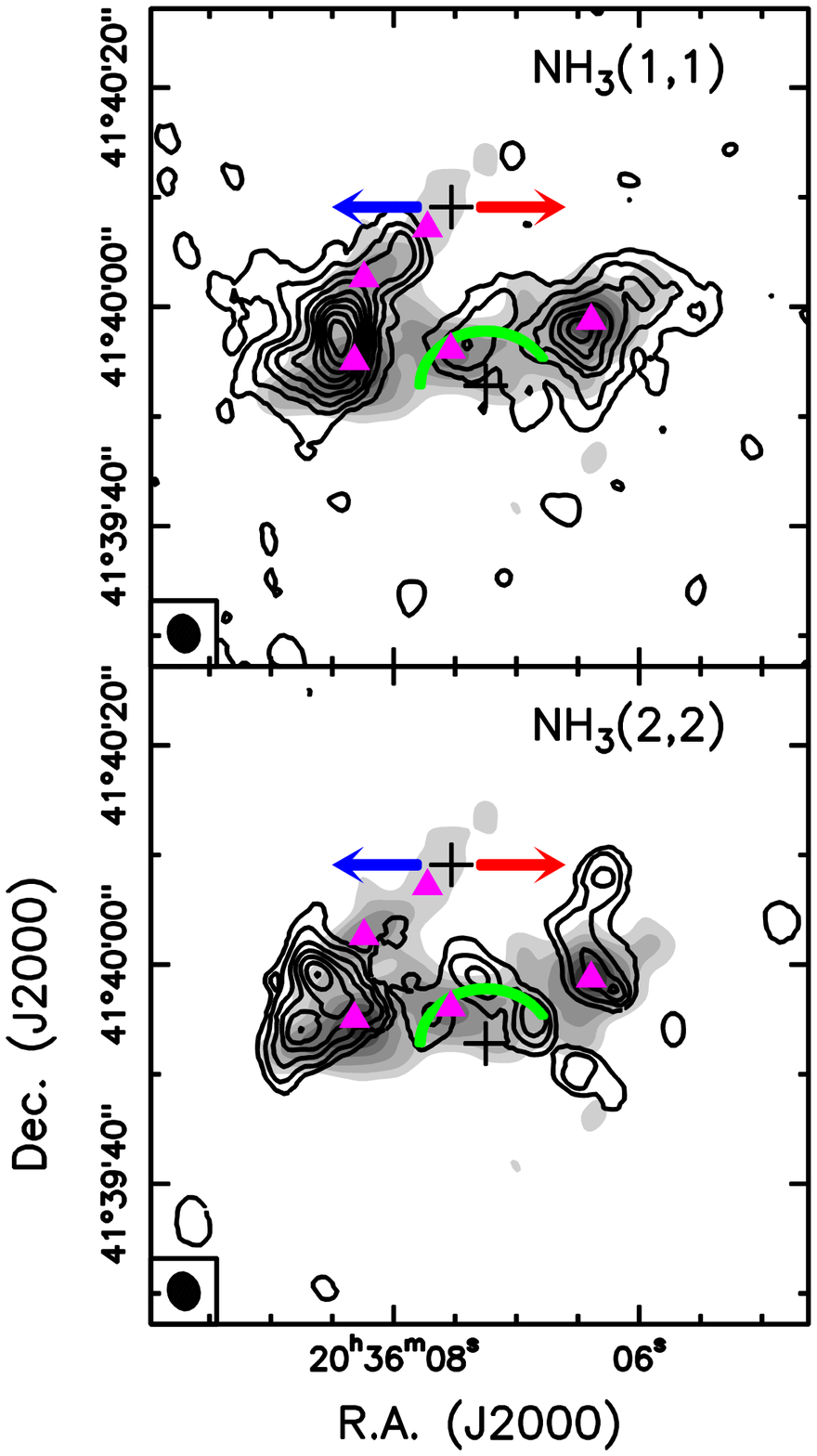}}
\caption{Top panel: integrated emission map of the ammonia (1,1) inversion transition 
(contours) observed with the VLA towards \ii , superimposed on the 3~mm 
continuum (CARMA, grey-scale). First contour level and step are the 10$\%$ of the 
peak (224~Jy~beam$^{-1}$~m~s$^{-1}$). 
The ellipse at bottom-left represents the VLA synthesised beam
(4\pas 20 $\times$ 3\pas 19, at P.A.~$\sim 11$\degr ).
Purple triangles pinpoint the 3~mm continuum peaks
(see Fig.~\ref{fig_cont}).
All other symbols (crosses, arrows and 
semiellipse) are the same as in Fig.~\ref{fig_cont}.
\newline
Bottom panel: same as top panel for the (2,2) inversion transition. 
First contour level and step are the 15$\%$ of the 
peak (57~Jy~beam$^{-1}$~m~s$^{-1}$). The synthesised beam
in this case is 4\pas 25 $\times$ 3\pas 19, at P.A.~$\sim 13$\degr .
}
\label{fig_nh3}
\end{figure}

The integrated intensity maps of \AMM\ (1,1) and (2,2) are presented
in Fig.~\ref{fig_nh3}. In the (1,1) line, the emission resembles
that seen with CARMA in the 3~mm continuum. Four main peaks are
detected,  which roughly correspond to those detected in
the 3~mm continuum.
On the other hand, none of the 1.3~mm continuum
peaks identified by Palau et al.~(\citeyear{palau07b}) exactly
coincides with any of the \AMM\ (1,1) peaks, despite the similar
angular resolution.
The \AMM\ (1,1) emission reveals two main clouds, one to the east
and the other to the west of IRS~3. The eastern cloud includes
MMD and MME and appears elongated
in the southeast-northwest direction, similar to the \H\ eastern cloud
and to the 3~mm continuum emission.
The western cloud has a condensation near MMB (near IRS~3)
and another condensation near MMA. The overall emission in the western
cloud is elongated in the east-west direction.
We stress that \AMM\ (1,1) emission is marginally detected also towards IRS 3
and the cavity driven by it, as for the 3~mm continuum.
On the other hand, the millimeter continuum sources associated
with IRS 1 as well as the \CO\ outflow lobes are not detected in ammonia.
The \AMM\ (2,2) transition is clearly detected towards the eastern side of MME,
where 2 peaks are resolved. The \AMM (2,2) line emission around IRS~3 
and towards the west is clumpy, peaking towards MMA
and near MMB and IRS3-SW.

From a comparison of the N-bearing to the C-bearing species, both \cdh\ and \cthd\
highlight clearly the cavity associated with IRS 3
(see top panels in Fig.~\ref{fig_CARMA_tot}),
while \nhdd\ and \AMM\ trace emission extending east-west
passing through IRS 3 (see Fig.~\ref{fig_nh3} and bottom panels of
Fig.~\ref{fig_CARMA_tot})
more similar to that seen in the 3~mm continuum map.
Thus, there seems to be a chemical dichotomy
in \ii\ among Carbon- and Nitrogen-bearing molecules. The exception
is represented by \H , detected away from IRS 3. This molecule appears to
trace the part of the cloud less disrupted by the expanding cavity.

\begin{figure*}
\centerline{\includegraphics[angle=-90,width=16cm]{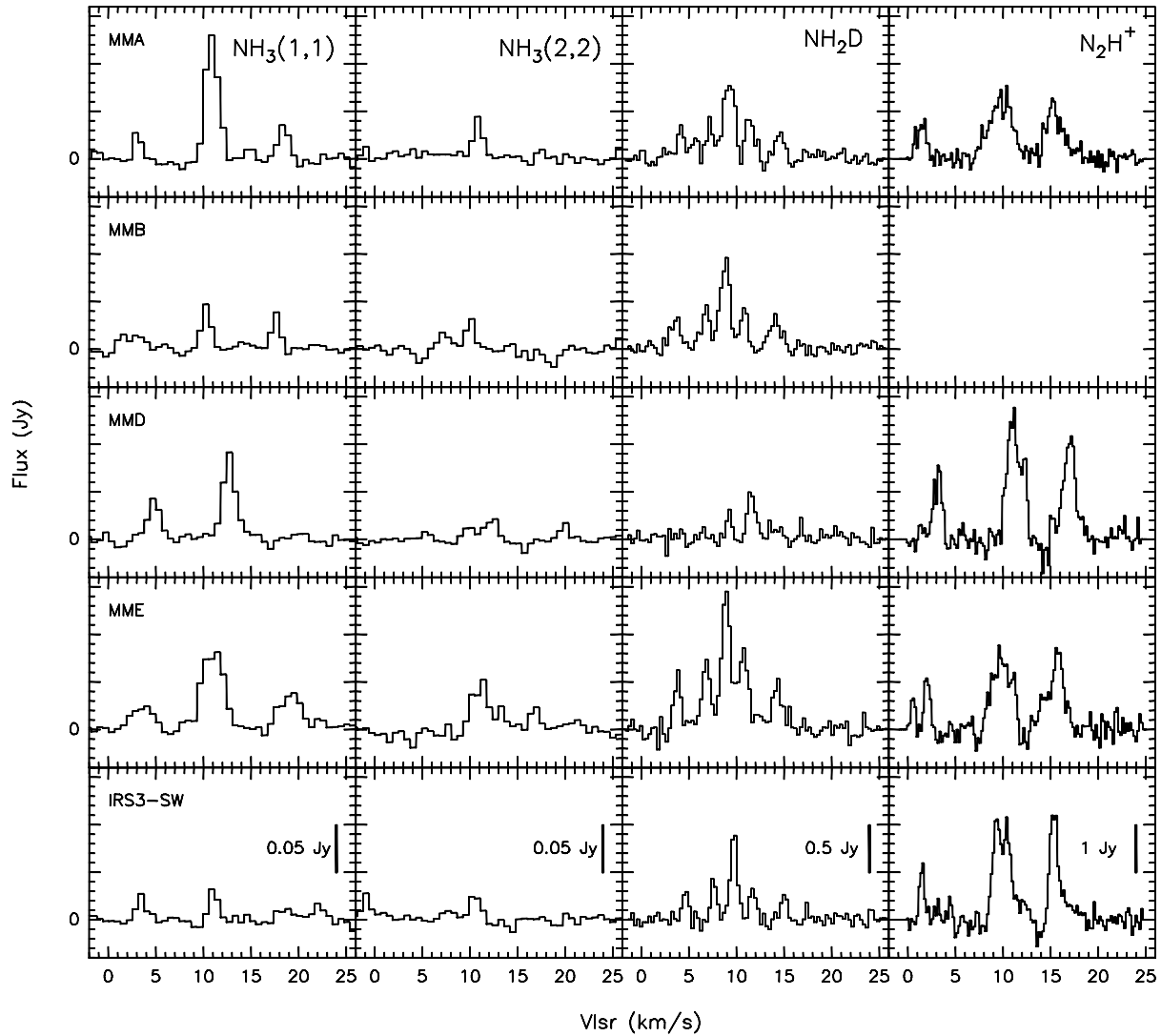}}
\caption{Spectra of the detected transitions of \AMM , \nhdd\ and \H\ at the position 
of the 3~mm continuum condensations (except MMC, undetected in all the lines) 
and the molecular condensation IRS3-SW.
The polygon used to extract the spectra corresponds to: the 5$\sigma$ rms
contour of the 3 mm continuum image for MMA and MME; the 3$\sigma$ rms 
contour of the 3~mm continuum image for MMB and MMD; the 5$\sigma$ rms contour
of the integrated emission of \nhdd\ (1--1) for IRS3-SW.}
\label{fig_specs}
\end{figure*}

The spectra of the \AMM (1,1) and \AMM (2,2) emission, integrated within the 5$\sigma$
contour polygon of the 3 mm continuum image (except for MMB and MMD, where we used the
3$\sigma$ contour) are shown in Fig.~\ref{fig_specs}. For IRS3-SW, we used the
5$\sigma$ rms contour of the \nhdd\ (1--1) line integrated emission (Fig.~\ref{fig_CARMA_tot},
bottom-right panel).
To compare the different molecular species, in Fig.~\ref{fig_specs} we also show the 
integrated spectra of \nhdd\ and \H\ extracted using the same polygons. 
We do not show the spectra towards MMC because this core is undetected in \nhdd\ and
\AMM\ (2,2), and marginally detected in the other lines.
Among all the \AMM\ (1,1) spectra, MMA shows the strongest emission, and 
MME shows the broadest lines, of up to 2.2 \kms . Such a large line broadening in MME could be due to a double
velocity component, as suggested by the \H\ spectrum which has 10 times better
spectral resolution. Concerning MMB, MMD, and IRS3-SW \AMM\ (1,1) spectra, it is
striking the anomaly seen in the inner satellite hyperfine lines, with one inner
satellite clearly detected above $5\sigma$ and the other inner satellite remaining
undetected. The anomaly for the non-LTE case due to hyperfine selective photon
trapping affects only the outer satellites (red stronger than blue, Stutzki \&
Winnewisser~\citeyear{sew}), allowing us to rule out this possibility in I20343. Rather,
anomalies of one inner satellite being stronger than the other have been observed
in several works (Lee et al. 2002, Longmore et al.~\citeyear{longmore}, Purcell et
al.~\citeyear{purcell}) and explained as being due to systematic motions, following the theoretical 
work of Park (\citeyear{park}). Park (\citeyear{park}) shows that if the core is contracting the inner blue
satellite should be stronger than the inner red satellite for a systematic motion
in the range of 0.4--1~\kms, and for a range of H$_2$ number densities and \AMM\ column
densities which are consistent with those derived by us (as we will show
in Sect.~\ref{res_cont}). This is the
case of MMD and IRS3-SW. On the other hand, if the core is undergoing expanding
motions, the prediction is that the inner red satellite should be stronger than the
inner blue satellite, as seen for the case of MMB. Thus, it seems that for these
three clumps the \AMM (1,1) anomalous intensity of the hyperfine components 
is consistent with contracting/expanding motions.

The MMA \AMM (1,1) spectrum is a very special case, as the inner satellites are detected at
an intensity smaller than that expected for LTE conditions (maximum
main-to-satellite ratio in LTE is 3.6, while the ratio for MMA is 4.3). This could
be explained if the opacity is high and the excitation temperature for the main line
and the satellites is different, with the satellites having a lower excitation temperature 
(and hence a smaller main beam temperature). A detailed discussion on the
temperature ratio between main line and satellites leads to a possible
non isothermal core made of two layers, with the external one being hotter
than the inner one. If this interpretation is correct, core MMA could be heated
externally, perhaps by IRS~3 and/or the infrared sources west of \ii\ (see right panel in
Fig.~\ref{fig_largefov} and Fig.~\ref{fig_CARMA_tot}).
We give details of this explanation in Appendix A.

We fitted the \AMM , \nhdd , and \H\ spectra in order to derive the physical parameters of
the gas traced by these molecules. To take into account the line hyperfine structure,
we followed the method described in the CLASS user manual\footnote{The CLASS program is
part of the GILDAS software, developed at the IRAM and the Observatoire 
de Grenoble, and is available at http://www.iram.fr/IRAMFR/GILDAS}. 
The \AMM\ (1,1), \nhdd\ (1--1) and \H\ (1--0) lines were fitted this way, while we
fitted the \AMM\ (2,2) lines with Gaussians. 
The derived fit parameters are reported in Table~\ref{tab_coldens}, except the line
velocities that will be extensively discussed in Sect.~\ref{moments}. These
parameters have been used to derive the molecular
column densities, the derivation of which will be described in Sect.~\ref{chemical}. 
To fit the hyperfine structure of MMB, MMD, and IRS3-SW, we used only the detected 
satellite and the main line, in order to obtain reliable opacities. 
For \AMM , in Table~\ref{tab_coldens} we list also the rotation temperature,
\Tr , obtained from the (2,2)-to-(1,1) intensity ratio following the
method outlined in Busquet et al.~(\citeyear{busquet09}),
which is based on the discussion presented in Ho \& Townes (1983).
These range from 13 K (in  MMD) to 23 K (in IRS3-SW).

\begin{table*}
\centering
\begin{minipage}{160mm}
\caption{\AMM , \H , and \nhdd\ line parameters for the 3~mm continuum cores (except MMC)
and the molecular core IRS3-SW. 
The parameters have been derived from the spectra
shown in Fig.~\ref{fig_specs} using the fitting
procedure described in Sect.~\ref{res_linesvla}. }
\label{tab_coldens}
\begin{tabular}{ccccccccccccc}
\hline \hline
Core & \multicolumn{4}{c}{ \AMM } & & \multicolumn{3}{c}{ \H } & & \multicolumn{3}{c}{ \nhdd }\\
         & $T_{\rm ex}$ & $\Delta v$ & $\tau$ $^{\it a}$ & \Tr\ & & $T_{\rm ex}$ & $\Delta v$ & $\tau$ $^{\it a}$  & & $T_{\rm ex}$ & $\Delta v$ & $\tau$ $^{\it a}$  \\
\cline{2-5} \cline{7-9} \cline{11-13} 
        & (K) & (\kms ) & & (K) & & (K) & (\kms ) & & & (K) & \kms\ &  \\
 MMA  & -- $^{\it b}$ & 1.52 & $> 1$ & 18(1)  & & 7 & 1.63 & 4.6  & & 17& 1.05 & 2.8  \\ 
MMB  & 9.7 & 1.03 & 3.5  & 20(7) & & -- & -- & -- & & 20 & 1.03 & 2.5  \\ 
MMD   & 21.5  & 1.21 & 1.4 & 13(2)  & & 14 & 0.9 & 5.2  & & 41 & 0.89 & 0.38  \\ 
MME & 19.8 & 2.2 & 1.5  & 18(3) & & 6 & 0.7 & 10 & & 26 & 1.01	& 3.2  \\ 
IRS3-SW & 7.9 & 1.03 & 3.2  & 23(7) & & 13 & 0.7 & 3.3  & &  23 & 0.72 & 1.8  \\ 
 \hline
\end{tabular}
\begin{flushleft}
$^{\it a}$ = Total opacity of the line; \\
$^{\it b}$ = in this core the excitation temperature for the main and the
 satellite lines of the \AMM\ (1,1) transition is probably different. 
See Appendix A for the detailed discussion on the derivation of the physical parameters from \AMM . \\
\end{flushleft}
\end{minipage}
\end{table*}

\begin{figure}
\centerline{\includegraphics[angle=0,width=9cm]{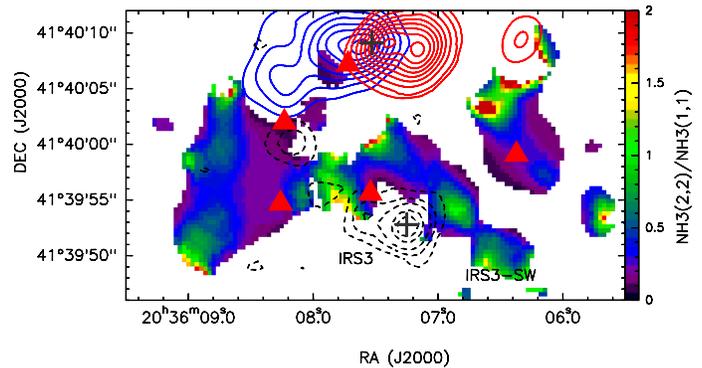}}
\caption{Colorscale: ratio of the integrated 
\AMM\ emission (2,2)/(1,1). Red triangles indicate the 3~mm continuum
peaks, which correspond to (from east to west) MME, MMD, MMC, MMB and MMA. 
The black-dashed contours indicate the VLA 1.3~cm emission (2.5, 5, 7.5, 12, 15 times 
50 $\mu$Jy beam$^{-1}$).
The blue/red solid contours depict the high-velocity 
CO\,(2--1) blueshifted/redshifted emission (Palau et al. \citeyear{palau07b}).
Blue contours range from 10 to 99\% in steps of 12\% of the peak intensity 
(21.335 Jy beam$^{-1}$ km s$^{-1}$); red contours are 6, 12 to 99\%  in steps of 
12\% of the peak intensity (41.781 Jy beam$^{-1}$ km s$^{-1}$). 
}
\label{fig_temp}
\end{figure}

In order to obtain a first approach to temperature variations of the dense gas across
I20343, we computed the ratio of the integrated \AMM\ (2,2)/(1,1), which has been
shown to be a reasonable approach to the kinetic temperature (e.g., Torrelles et al.
1993; Zhang et al. 2002). The result is shown in Fig.~\ref{fig_temp}.
As can be seen from the figure, the ratio is largest towards three main
regions: near IRS~3 (MMB, and IRS~3-SW), to the north of MMA, and
to the eastern side of MME. Thus, the spectra toward MMA, which can be explained
with external heating, is consistent with the 22/11 ratio map which reveals that the
heating comes most likely from the north of MMA. Interestingly, the north of MMA is
spatially coinciding with the redshifted CO outflow lobe driven by IRS~1, suggesting
that the outflow and the dense gas are interacting (Fig~\ref{fig_temp}). Most
intriguing is the high ratio seen at the eastern side of MME, which extends all
along the north-south direction and is seen westwards of the H$_2$ extended
emission reported by Kumar et al. (2002). Such a spatial coincidence is suggestive
of a possible relation between the high 22/11 ratio and the H$_2$ emission.
Finally, the high 22/11 ratio near IRS~3 could be indicative of direct heating by the
UC \HII\ region associated with IRS~3.

%

\subsection{Kinematics}
\label{moments}

To inspect the velocity field, we have extracted from the
interferometric channel maps spectra
of the molecular transitions detected on grids 
with regular spacings (1.5\arcsec\ $\times$ 1.5\arcsec\ for \cdh , 
2.5\arcsec\ $\times$ 2.5\arcsec\ for \cthd , 1\arcsec\ $\times$ 1\arcsec\ for \H ,
1.6\arcsec\ $\times$ 1.5\arcsec\ for \nhdd\ and \AMM ). 
The spectra extracted have been fitted following different methods:
for \cdh\ (1--0), \cthd\ (2--1) and \AMM\ (2,2) we assumed Gaussian lines, while 
for \H\ (1--0), \nhdd\ (1--1) and \AMM\ (1,1) we fitted the lines 
following the method described in Sect.~\ref{res_linesvla} to take into
account their hyperfine structure. 

\subsubsection{Line peak velocities}

\begin{figure*}
\centerline{\includegraphics[angle=0,width=16cm]{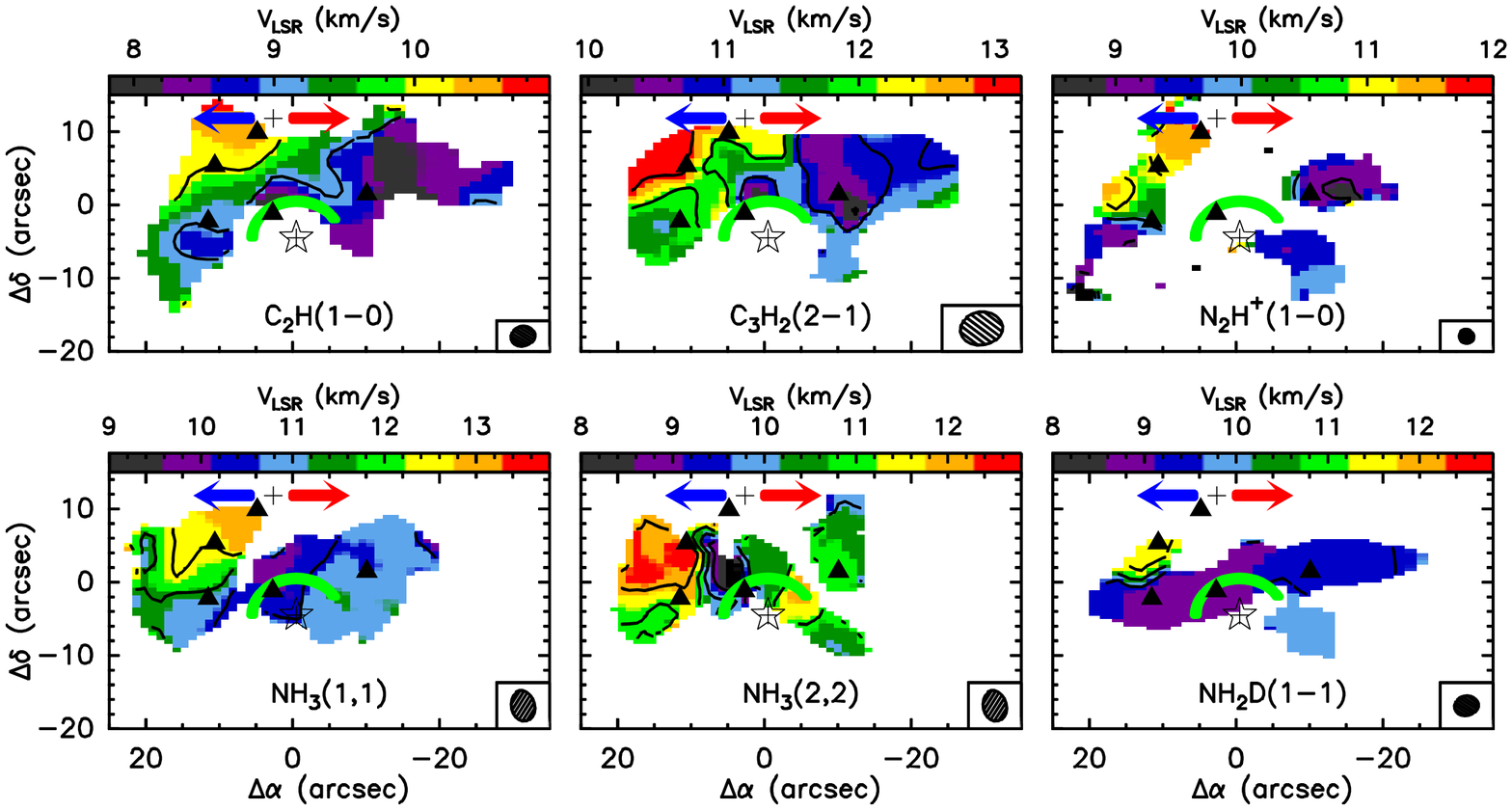}}
\caption{Maps of the line peak velocity for all detected molecular lines.
Top panels show (from left to right): \cdh\ (1--0), \cthd\ (2--1) and \H\ (1--0). 
Contours are in steps of 1 \kms , and range from: 
8 to 11 \kms\ for \cdh ;  
10.5 to 13.5 for \cthd ; 9 to 12 \kms\ for \H . 
Bottom panels show (from left to right): \AMM\ (1,1) \AMM\ (2,2) and
\nhdd\ (1--1). Contours are in steps of 1 \kms , and range from: 
9.5 to 13.5 \kms\ for \AMM\ (1,1); 9 to 13 \kms\ for \AMM\ (2,2); 
8.5 to 12.5 \kms\ for \nhdd . In each panel, the ellipse 
in the bottom-right corner represents the CARMA or VLA synthesised beam.
The black triangles pinpoint the 3~mm continuum peaks (Sect.~\ref{continuum}).
The position of IRS 1 and IRS 3, as well as the associated outflow and fan-shaped
H$_2$ emission, are depicted in each panel as in Fig.~\ref{fig_cont}.}
\label{fig_vel}
\end{figure*}

Fig.~\ref{fig_vel} shows maps obtained from the line peak velocities. In all
tracers the radial velocities
are predominantly blue-shifted to the west and red-shifted to the north-east. 
The east-west velocity gradient is not uniform and
suggests a possible torsion of the gas. Interestingly, the inclination of
this gradient with respect to the line 
of sight is opposite to that of the outflow associated with IRS~1, since the 
blue lobe of the outflow is located on the side where the dense gas is red-shifted, 
and vice-versa. In general, all tracers with
emission near IRS~3 (in MMB) show that the gas is blueshifted at this
position.

\subsubsection{Line widths}

\begin{figure*}
\centerline{\includegraphics[angle=0,width=16cm]{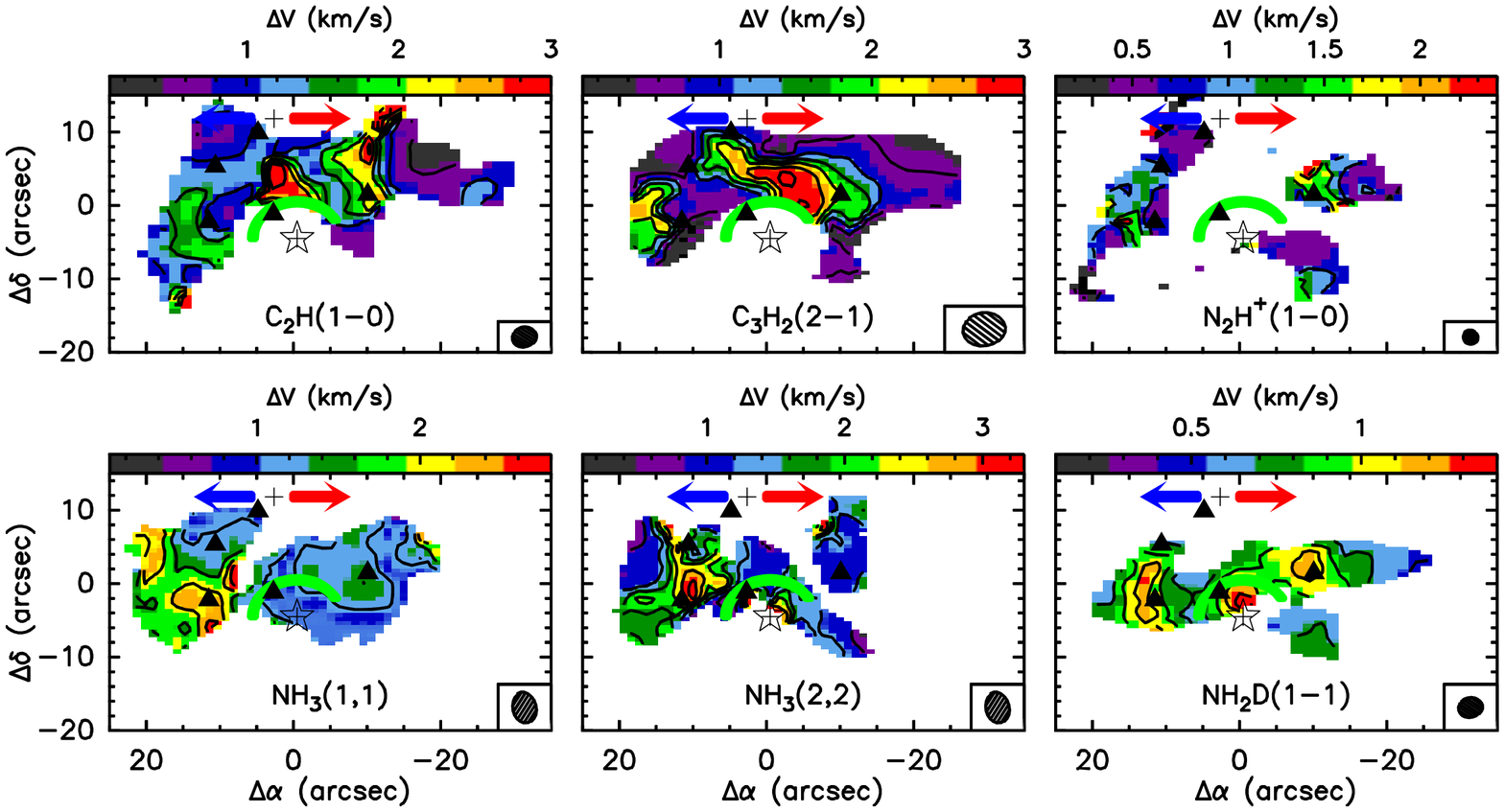}}
\caption{Same as Fig.~\ref{fig_vel} for the line widths. Contours
are in steps of 0.5 \kms\ and range from: 0.5 to 3 \kms\ for \cdh ; 
0.6 to 3.1 \kms\ for \cthd ; 0.4 to 2.4 \kms\ for \H ; 0.6 to 2.6 \kms\ for \AMM\ (1,1); 
0.8 to 3.2 \kms\ for \AMM\ (2,2) ; 0.3 to 1.3 \kms\ for \nhdd\ (1--1).}
\label{fig_width}
\end{figure*}

Maps of the line widths are presented in Fig.~\ref{fig_width}. The
measured line broadenings are generally a factor $>3$
higher than the thermal broadening, expected to be of the order of
$\sim 0.1-0.3$ \kms , indicating that the gas kinematics is largely dominated by
non-thermal motions. This finding confirms previous studies in similar 
intermediate- to high-mass protoclusters (e.g.~Palau et al.~\citeyear{palau07a}, 
Fontani et al.~\citeyear{fonta09}, Busquet et al.~\citeyear{busquet}),
and represents one of the most important differences between
these dense cores and those observed in low-mass star forming regions,
where the line widths are dominated by thermal broadening, even
in clustered environments (e.g.~Kirk et al.~\citeyear{kirk}, 
Walsh et al.~\citeyear{walsh}, Bourke et al.~\citeyear{bourke}).


In summary we highlight three regions where the gas is more 
turbulent (Fig.~\ref{fig_width}): (i) around IRS~3, especially in between IRS 1 and
IRS 3 (see the \cdh\ and \cthd\ line widths in Fig.~\ref{fig_width});
(ii) north of MMA, near the red lobe; (iii) towards MMA.
The turbulence enhancement around IRS~3 is easily explained by
the expading cavity, while in MMA and north of it could rather be
due to the red-lobe of the outflow.
On the other hand, we stress that towards MME and in between IRS~1 
and IRS~3 some spectra show a double-velocity component 
(see e.g. the isolated hyperfine
component in the \H\ (1--0) line towards MME in Fig.~\ref{fig_specs}). 
These are especially evident in the \H , \AMM\ 
and \nhdd\ spectra, but the second component can be seen also in
some spectra of \cdh\ and \cthd\ close to the cavity. 
Therefore the
large broadening above IRS~3 and in MME could just be due 
to the superposition of two unresolved velocity components.

Interestingly, for \cdh\ and \H\ the line broadening is relatively 
small close to the blue lobe of the outflow associated with IRS~1 
and larger besides the red lobe, while the opposite seems to occur
for the \cthd\ (2--1) line (top-middle panel in Fig.~\ref{fig_width}). 
This could indicate a selective influence of the flow on the
different molecules in the surrounding material.
However, we stress that the \cthd\ emission is very faint at the borders 
of the region plotted in Fig.~\ref{fig_width}, where the fit results are affected
by large relative errors.


\subsubsection{Position-velocity plots}

\begin{figure}
\centerline{\includegraphics[angle=0,width=8cm]{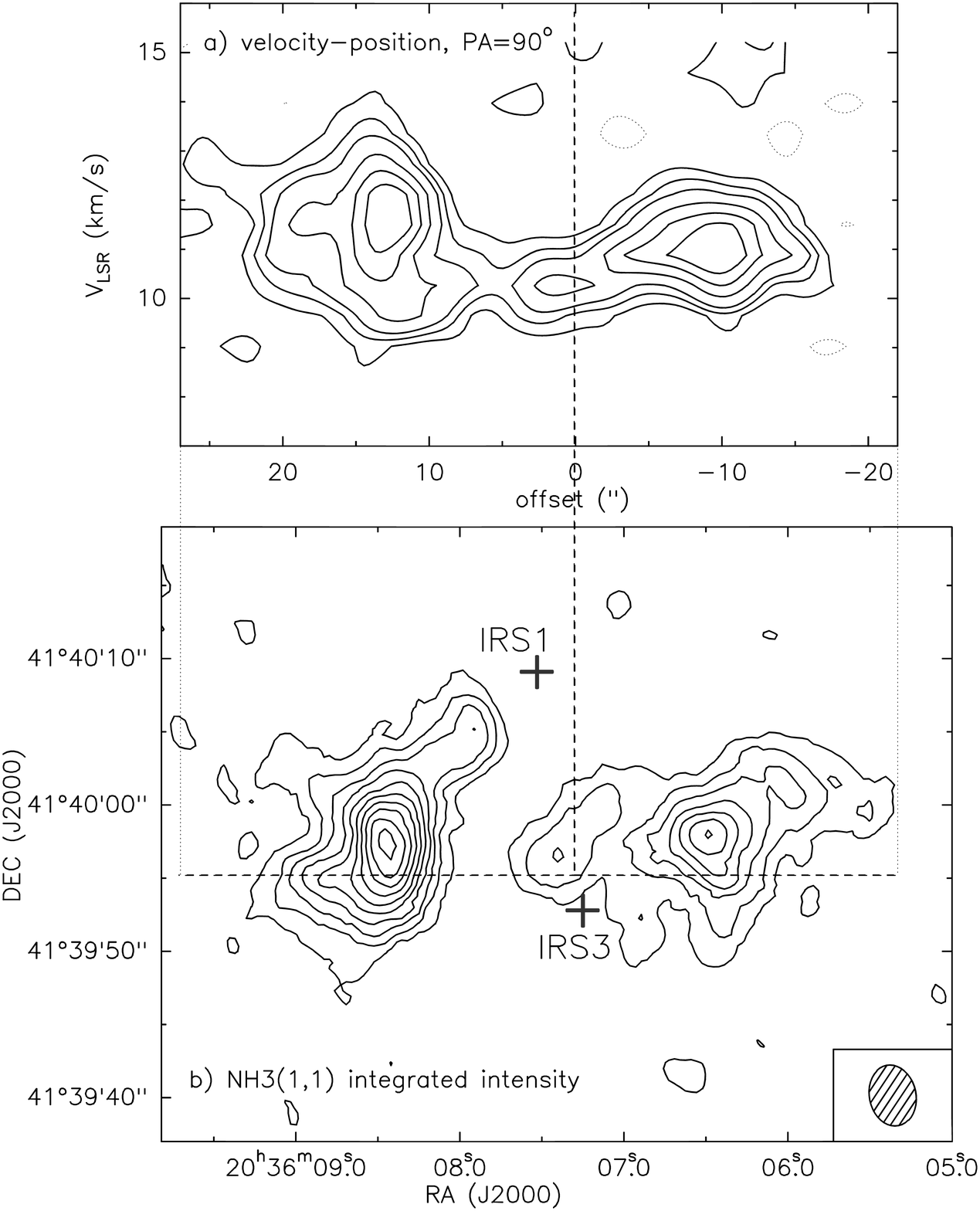}}
\caption{a) \AMM\ (1,1) position-velocity plot centred at offset (0,$-2$\asec ) with 
respect to the CARMA phase centre and along a cut in the east-west 
direction. The \AMM\ (1,1) data cube was convolved with a beam of 
5\asec$\times 2$\asec\ and PA=0\degr\ to increase signal-to-noise.  Contours 
are: $-3$, 3, 6, 9, 15, 21, and 27 times 0.001 Jy beam$^{-1}$.
b) \AMM\ (1,1) integrated emission map (same contours as in Fig.~5) with the cut of the 
position-velocity plot of panel "a" indicated by a dashed line. Contours start at 12\% of 
the peak intensity,  224 Jy beam$^{-1}$ m s$^{-1}$, and increase in steps of 10\%. 
The VLA synthesised beam is shown in the bottom-right corner.}
\label{fig_pv}
\end{figure}

In order to further study the velocity field of the \AMM\ emission near IRS~3, we
computed position-velocity plots for the \AMM\ (1,1) line in the east-west direction, 
and centred at offset 0\arcsec ,$-2$\arcsec\ with respect to the CARMA phase centre. 
The final plot is shown in panel "a" of Fig.~\ref{fig_pv}. The \AMM\ (1,1) emission in the 
position-velocity plot shows two main peaks, one corresponding to the eastern 
cloud and the other corresponding to the western cloud, and both peaks are linked 
through fainter emission which overall shows a U-like structure. 
Such a feature resembles the shape predicted by the model of 
Arce et al.~(\citeyear{arce}) for an expanding bubble.  In this model, an 
expanding shell would appear as a ring in the (p-v) plots (see their Fig.~5), 
while we only see the blue-shifted half of it. However, a U-like feature can be
explained if the source driving the bubble is slightly displaced (behind 
the bubble) with respect to the surrounding molecular gas, so 
that we mainly see the gas which is moving towards us.
In fact, the tails of the U-like feature are found, as expected, 
at approximately the systemic velocity.
This suggests that IRS~3 may be pushing the surrounding 
dense material away (either through its winds/radiation or through the
associated \HII\ region), with an expansion velocity
(difference between the 'tip' of the U-like feature and the 'tails')
of about 2 \kms\ (see panel "a" in Fig.~\ref{fig_pv}). A similar expanding shell was recently
found around an infrared-source at the centre of a region devoid of
gas emission in the intermediate- to high-mass protocluster IRAS 05345+3157
(Fontani et al.~\citeyear{fonta12}). 


\subsection{Physical parameters from 3~mm and 1.3~cm continuum emission}
\label{res_cont}

\subsubsection{3~mm}

We have identified 5 main condensations in the 3~mm continuum emission 
map (see Sect.~\ref{continuum}). Their peak positions are given in Cols. 2 and 3 
of Table~\ref{cont_tab}. We also list the angular ($\theta $) and linear 
($D$) diameters (Cols.~6 and 7 in Table~\ref{cont_tab}) computed assuming the 
sources are Gaussians, and deconvolving the 
contour at half maximum with a Gaussian beam with HPBW corresponding to the 
geometric mean of the minor and major axes of the CARMA synthesised beam
(see Table~\ref{obs_par}). 
Because the contours at half maximum were blended, at the edge between two cores
we decided to separate the emission arising from the different condensations
identifying the peaks and considering the first unblended contour. 
The same criterion was applied to derive the integrated flux density, $F_{\nu}$, 
given in Col.~5 of Table~\ref{cont_tab}.

From $F_{\nu}$, we have computed the mass of the condensations under the
assumptions that the dust millimeter-continuum emission is optically thin, 
and that the dust temperature equals the gas kinetic temperature. This latter 
hypothesis implies coupling between gas
and dust, which is a reasonable assumption for H$_2$ volume densities above 
10$^{5}$ \cmc . Under these assumptions, the gas mass can be derived from the
formula:
\begin{equation}
M=\frac{F_{\nu} d^{2}}{B_{\nu}(T) k_{\nu}}\,\,,
\label{eq_mass}
\end{equation}
where $d$ is the source distance, $B_{\nu}(T)$ is the Planck function at dust temperature $T$, 
and $k_{\nu}$ is the dust opacity per 
unit dust mass. For this latter, we extrapolated the value at 230 GHz given by
Kramer et al.~(\citeyear{kramer}), $k_{\rm 230}=0.005$ cm$^2$ g$^{-1}$
(which assumes a gas-to-dust ratio of 100),
through the power-law $k_{\nu}=k_{\rm 230}{[\rm \nu (GHz) /230]^{\beta}}$.
We have assumed $\beta =2$, which is a typical value derived for dusty 
envelopes of massive (proto)stellar objects (e.g.~Molinari et al. 2000, Hill et al.~\citeyear{hill}).
As gas temperature, we have taken the kinetic temperature
obtained by extrapolating the rotation temperature
derived from the ammonia $(2,2)/(1,1)$ line ratio for each core
(see Sect.~\ref{res_linesvla} and Table~\ref{tab_coldens}) 
following the empirical approximation 
method described in Tafalla et al.~(\citeyear{tafalla}).
The kinetic temperatures derived this way are listed in Col.~4 of Table~\ref{cont_tab},
and are in between 14 and 25~K, higher (a factor 1.5 -- 2)
than the values measured typically in low-mass clustered starless cores 
(e.g.~Andr\'e et al.~\citeyear{andre}, Foster et al.~\citeyear{foster}) which
are around 10--13 K.
For MMC, undetected in \AMM\ (2,2), we decided to give a range
of masses computed in the temperature interval 15--30~K.

The resulting masses are listed in Col.~8 of Table~\ref{cont_tab}. All fragments
have masses consistent with intermediate-to high-mass embedded objects.
The most massive one is MME (23~\solm ).
By assuming spherical and homogeneous cores, we have derived the
average H$_2$ volume and column densities. The average volume and column densities 
(given in Cols.~9 and 10 of Table~\ref{cont_tab}) are of the 
order of $10^{6-7}$\cmc\ and $10^{23-24}$\cmq . Such high column
densities are consistent with
being the birthplaces of intermediate- and/or high-mass objects
(e.g.~Krumholz \& McKee~\citeyear{kem}).


\begin{table*}
\centering
\begin{minipage}{160mm}
\caption{Peak position, angular and linear diameter, integrated flux density, mass, H$_2$ volume 
and column density of the 3~millimeter condensations detected with CARMA. The masses are computed 
for $\beta = 2$, and assuming the temperatures derived from ammonia. The H$_2$ volume and
column densities are calculated assuming a spherical source with diameter equal to the deconvolved
level at half maximum. $F_{\nu}$ has been obtained by integrating the continuum flux density
inside the 3$\sigma$ contour level. }
\label{cont_tab}
\begin{tabular}{cccccccccc}
\hline \hline
Core & \multicolumn{2}{c}{Peak position} & \Tk\ $^{\it a}$ & $F_{\nu}$ & \multicolumn{2}{c}{Diameter}  & $M_{\rm cont}$ & $n_{\rm H_2}$  & $N({\rm H_2})$  \\
\cline{2-3} \cline{6-7} 
       & R.A.(J2000) & Dec.(J2000)  & & & $\theta_{\rm s}$  & $D$  &  &  & \\
       & 20$^h$36$^m$  & & (K)  & (mJy) & (\asec)& (pc) & (M$_{\odot}$) & ($\times 10^6$ \cmc ) & ($\times 10^{23}$ \cmq ) \\
       MMA   & 06$^s$.37 &  $+41\degr$39$^{\prime}$59\pas 1 & 22$^{\it b}$ &  7.4   &    7.3  &   0.049  &    21 &  6.7 &  10.3  \\
MMB  & 07$^s$.54 & $+41\degr$39$^{\prime}$55\pas 6       &  25 & 4.6   &    6.8 &   0.046   &   11 & 4.5  & 6.4  \\
MMC  & 07$^s$.72 & $+41\degr$40$^{\prime}$07\pas2      & -- &  1.9   & 5.5  & 0.037 &  8$^{\it c}$,4$^{\it d}$ & 6.2$^{\it c}$,2.9$^{\it d}$ & 7.1$^{\it c}$,3.3$^{\it d}$   \\
MMD   & 08$^s$.23 & $+41\degr$40$^{\prime}$02\pas 0    & 14 &   3.7    &   6.3  &  0.043    &  17 &  8.7 &  11.4 \\
MME & 08$^s$.26 & $+41\degr$39$^{\prime}$54\pas 7       &  22 & 8.3  &   8.0 &   0.054   &    23 &  5.7 &  9.6  \\
 \hline
\end{tabular}
\begin{flushleft}
$^{\it a}$ = Derived from \Tr\ as explained in Sect.~\ref{res_cont}; \\
$^{\it b}$ = see Appendix A for the derivation of \Tk\ from ammonia for this core; \\
$^{\it c}$ = assuming a temperature of 15~K; \\
$^{\it d}$ = assuming a temperature of 30~K. \\
\end{flushleft}
\end{minipage}
\end{table*}

\begin{figure}
\centerline{\includegraphics[angle=270,width=10cm]{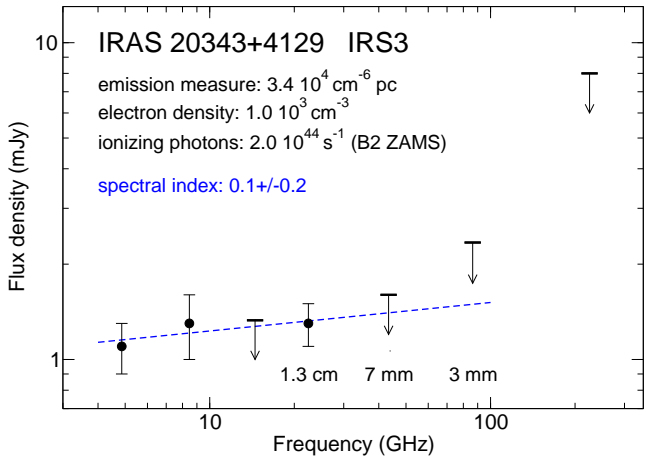}}
\caption{Spectral energy distribution of IRS~3 in the cm/mm range.
The physical parameters have been derived assuming the emission at 1.3~cm 
comes from an optically thin \HII\ region.
}
\label{fig_contcm}
\end{figure}

\subsubsection{1.3~cm}

In Sect.~\ref{continuum} we showed that the centimeter emission is dominated by one single 
source associated with IRS~3 and extending towards the east. 
In addition, a secondary source near MMD was also identified. 
A 3.6 and 6~cm source associated with IRS~3 is already reported by 
Miralles et al.~(\citeyear{miralles}) and Carral et al. (1999). 
However, from these two measurements only, and taking into 
account the uncertainties, the spectral index of the centimeter source 
associated with IRS~3 could not be well determined (e.g., Palau et al. \citeyear{palau07b}). 
Our new measurement at 1.3~cm allows to better constrain the spectral index of the 
source to $0.1\pm0.2$ (Fig.~\ref{fig_contcm}). 
Such a spectral index is consistent with optically thin free-free emission
favoring the interpretation that the centimetre emission comes from an \HII\
region rather than a thermal radio jet (which typically have steeper spectral
indices). We calculated the physical parameters of the ionised region at
1.3~cm assuming the emission is optically thin, and obtained an emission
measure of $3.4 \times10^4$~cm$^{-6}$~pc, characteristic of UC \HII\
regions, and a flux of ionising photons of $2 \times10^{44}$~s$^{-1}$,
consistent with the \HII\ region being ionised by an early-type B2 star
(Panagia~1973).
Interestingly, the extension of such an \UC\ region 
towards the east is similar (although larger in size) to the extension seen 
in [NeII] emission by Campbell et al. (\citeyear{campbell}), and which 
is interpreted as due to the expansion of the ionised gas and disruption 
of the natal envelope.  In this context, the secondary centimetre peak near 
MMD could be related as well to the expanding ionised gas. We estimated the 
possible contribution of free-free emission to the flux measured at 3~mm for MMB, 
and is of $\sim0.3$~mJy (4$\sigma$ of 1.3~cm observations, and using the spectral index of 0.1), 
out of 4.6 mJy, or 6\%. 
Thus, thermal dust emission is the main contribution to 
the 3 mm continuum emission in MMB.
Finally, we estimated an upper limit for the 3 mm emission associated with IRS~3
of 2.3~mJy,
measured as the 3 mm flux density 
inside the 4$\sigma$ contour of the centimeter 
emission (Fig.~\ref{fig_contcm}), and we cannot rule out the possibility of the \UC\ 
region being still associated with remnant natal dust, although a projection effect
could be also possible.


\section{Discussion}
\label{discu}

\subsection{Column densities of the PDR tracers \cdh\ and \cthd }
\label{dis_cchc3h2}

The two carbon-bearing species \cdh\ and \cthd\ are among
the most abundant simple carbon-chain molecules detected in the interstellar
medium, and are believed to be good tracers of PDRs
(Lucas \& Liszt~\citeyear{lel}, Pety et al.~\citeyear{pety}, Gerin et al.~\citeyear{gerin}).
\cdh\ is formed either from photodissociation of acetylene
(C$_2$H$_2$) followed by dissociative recombination of C$_2$H$_2^+$
(Mul \& McGowan~\citeyear{mem}) or through neutral-neutral reaction
between C and CH$_2$ in hot gas (Sakai et al.~\citeyear{sakai}).
\cthd\ is believed to be formed by dissociative recombination of
{\it c-}C$_3$H$_3^+$. Both species benefit
from the presence of atomic carbon not locked in CO, and a good
correlation between the two tracers has been found at the illuminated 
surface of the Horsehead nebula (Pety et al.~\citeyear{pety}, 
Gerin et al.~\citeyear{gerin09}), as well as in both diffuse and translucent 
clouds (Lucas \& Liszt~\citeyear{lel}, Gerin et al.~\citeyear{gerin}).

We have investigated the relation among the two species in \ii .
For this purpose, we have extracted the spectra of \cdh\ (1--0) and \cthd\ (2--1) 
on a grid of spacing 2.5\arcsec\ $\times$ 2.5\arcsec (roughly half of the CARMA 
synthesised beam at the frequency of the \cthd\ (2--1) transition), 
and fitted the spectra with Gaussian lines.
Then, from the integrated intensity obtained from the fits, we have computed
the column densities assuming that both lines are optically thin. This assumption is mandatory 
because the opacity of the lines cannot be directly measured
(for the \cdh\ (1--0) line we observed only the main hyperfine component,
and we do not have isotopologues for \cthd ). 
However, given that the line profiles generally do not show effects due
to high optical depths, we are confident that the assumption is reasonable.
We used the general formula for optically thin transitions (compare to, 
e.g., Eq.~A.3 in Pillai et al.~\citeyear{pillai07}):
\begin{eqnarray}
\lefteqn{N_{\rm tot} = \frac{3 h}{8 \pi^3} 
  \frac{Q(T_{\rm ex})}{ S \mu^2} 
  \frac{W}{J_{\nu}(T_{\rm ex})-J_{\nu}(T_{\rm BG})} 
  \frac{{\rm e}^{\left(\frac{E_J}{kT_{\rm ex}}\right)}}{{\rm e}^{h \nu / k T_{ex}}-1} }
\label{eq_nj}
\end{eqnarray}
where: $E_J$ and $S$ are energy of the upper level and line strength, 
respectively, 
$W$ is the integrated intensity of the line, $Q(T_{\rm ex})$ is 
the partition function at the temperature $T_{\rm ex}$, 
$\nu$ the line rest frequency, $J_{\nu}(T_{\rm ex})$ and $J_{\nu}(T_{\rm BG})$ are 
the equivalent Rayleigh-Jeans temperature at frequency $\nu$
computed for the excitation and
background temperature ($T_{\rm BG}\sim 2.7~K$), respectively;
$\mu$ the molecule's dipole moment ($0.77$ Debye for \cdh\ and $3.27$ 
Debye for \cthd ).
For \cdh , in Eq.~(\ref{eq_nj}) $W$ has been obtained by multiplying
the integrated emission of the hyerfine component observed for its 
relative intensity (0.416).
As excitation temperature, we have assumed a reasonable
value of 20~K based on the excitation temperatures computed
for the other lines (see Table~\ref{tab_coldens}).
The values of $E_J$, $S \mu^2$ and $Q$ have been taken
from the Cologne Database for Molecular Spectroscopy 
(CDMS\footnote{http://www.astro.uni-koeln.de/cdms/}, 
M\"uller et al.~\citeyear{muller}).
For this latter, we have extrapolated the values tabulated
to an excitation temperature of 20~K. For \cthd , the {\it ortho-/para-}
ratio is included in the partition function.

\begin{figure}
\centerline{\includegraphics[angle=-90,width=8cm]{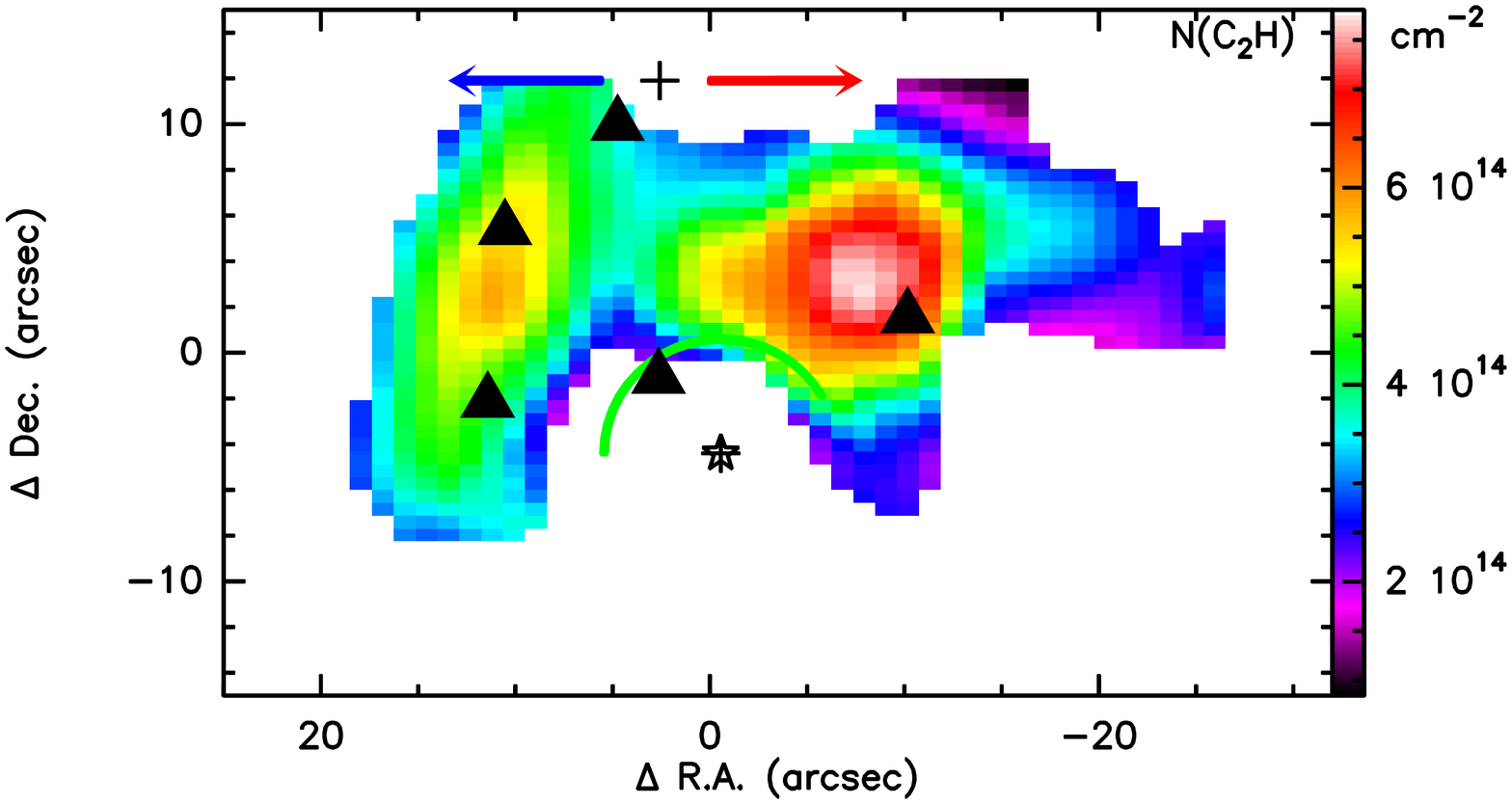}}
\centerline{\includegraphics[angle=-90,width=8cm]{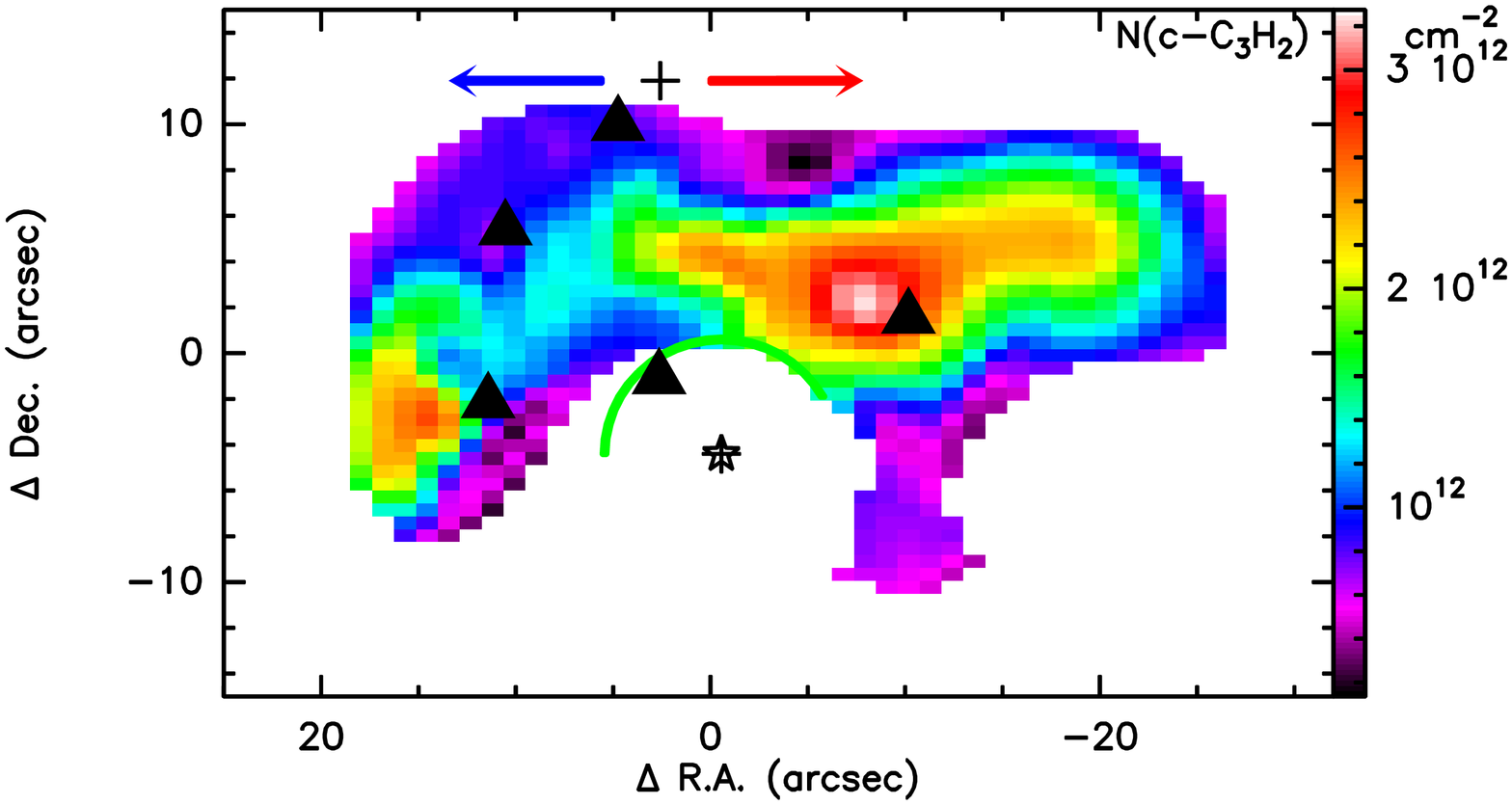}}
\centerline{\includegraphics[angle=-90,width=8cm]{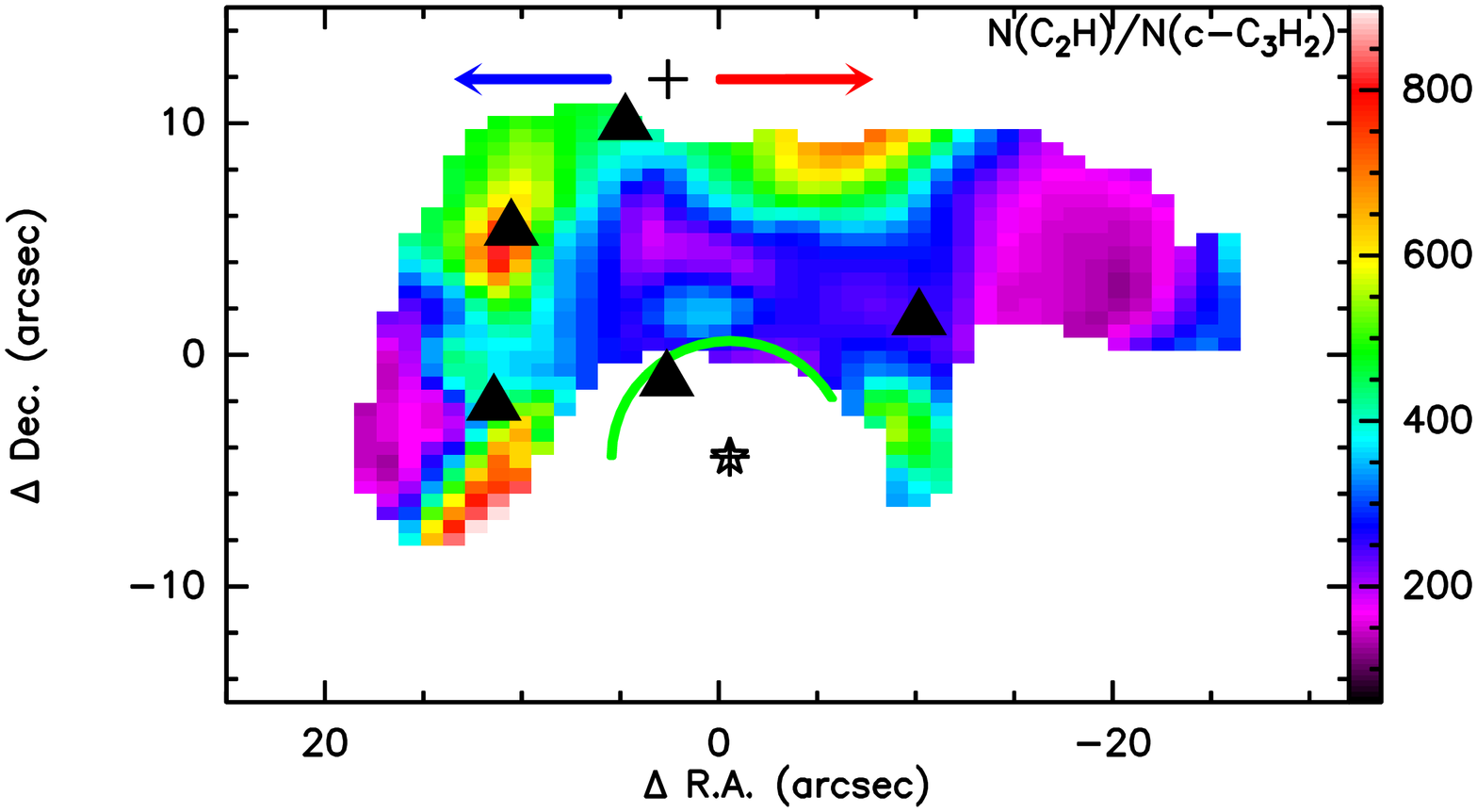}}
\caption{From top to bottom: Column density map of \cdh\ and
\cthd , and their ratio, derived from the CARMA observations. 
For the column density derivation of both species, an excitation temperature 
of 20~K has been adopted. The other symbols are the same as in
Figs.~\ref{fig_vel} and Fig.~\ref{fig_width}.}
\label{fig_coldens}
\end{figure}

The results are shown in Fig.~\ref{fig_coldens}. The column density
of \cdh\ is of the order of $10^{14}$ \cmq\ across the cloud (Fig.~\ref{fig_coldens}, top panel), while
that of \cthd\ is of the order of $10^{12}$ \cmq\ (Fig.~\ref{fig_coldens}, middle panel). The \cdh\ column 
densities are generally larger than those found by Gerin et al.~(\citeyear{gerin}), who
measured column densities of $10^{13}$ \cmq ,
while those of \cthd\ are more consistent. Their ratio is on average
of the order of 200 - 400 (Fig.~\ref{fig_coldens}, bottom panel), i.e.
one order of magnitude larger than the value 20 - 30 measured
in translucent clouds (Gerin et al.~\citeyear{gerin}), as well as in diffuse
high latitude clouds (Lucas \& Liszt~\citeyear{lel}) and in the
Horsehead nebula (Pety et al.~\citeyear{pety}).
On the other hand, the chemical models of PDRs in Gerin et al.~(\citeyear{gerin})
seem to be more consistent with our observational results 
rather than with theirs, because the models predict total
column densities consistent with our values for both species, 
and ratios of the order of 100 or even more (see their Table~5). 
Interestingly,
we find the largest
ratios (around 400--800) close to the outflow lobes and to the east and west of the
cavity walls, where the gas is probably less dense
because most disrupted.
Significant enhancement can be noticed also in the eastern clump, 
in between MMD and MME, where 1.3~cm emission is detected
(see Sect.~\ref{continuum} and Fig.~\ref{fig_cont}).
This would confirm strongly that both molecules are produced
in PDR regions, and that they are maybe tracing a low density envelope
in which the dense cores detected in \H\ and \AMM\ are embedded. 

\subsection{Chemical differentiation and nature of the 3~mm continuum cores}
\label{chemical}

Studies of intermediate- and high-mass star-forming regions
suggest that the relative abundance ratio \AMM -to-\H\ 
is an evolutionary indicator for dense cores (e.g.~Palau et al.~\citeyear{palau07a}, Fontani
et al.~\citeyear{fonta12}): cores with no signs of star formation typically have larger
\AMM -to-\H\  column density ratio than cores associated with active star
formation. Also, Fontani
et al.~(\citeyear{fonta08}) and Busquet et al.~(\citeyear{busquet}) have
measured that the deuterated fraction (i.e. the column density ratio
of a deuterated specied to that of the hydrogenated counterpart)
of \AMM\ and \H\ is of the order of 0.1 in pre--protostellar core candidates,
as high as in low-mass pre--stellar cores, while it is lower in more evolved
objects.
With this in mind, we have investigated the evolutionary stage of the millimeter cores
in \ii\ based on the column density ratios \AMM -to-\H\ and \nhdd -to-\AMM .
From the line parameters derived in Sect.~\ref{res_linesvla} (see
Table~\ref{tab_coldens}) we have
computed the total column densities of \AMM , \nhdd\ and \H\
from Eq. (A1) of Caselli et al.~(\citeyear{caselli02})
because all lines are optically thick (except \nhdd\ (1--1) in MMD,
but the opacity is well-constrained). As for the other parameters,
for a detailed discussion on the \AMM\ data of MMA 
see the Appendix A.

\begin{table*}
\caption{Total column densities of \AMM , \nhdd\ and \H\ for the 3~mm continuum cores, except MMC,
and the additional core IRS3-SW.}
\label{tab_coldensratio}
\begin{center}
\begin{tabular}{cccccc}
\hline 
core & $N$(\AMM ) & $N$(\H ) & $N$(\nhdd ) & $N$(\nhdd )-to-$N$(\AMM ) & $N$(\AMM )-to-$N$(\H ) \\
        & ($\times 10^{15}$ \cmq ) & ($\times 10^{13}$ \cmq ) & ($\times 10^{14}$ \cmq ) &  & \\
\hline
MMA & $> 2.41$\footnote{see Appendix A for details on the derivation of the \AMM\ physical parameters} & 7(2) & 2.2(0.2) & $< 0.09$ & $>34$ \\
MMB & 2.3(0.6) & -- & 2.6(0.3) & 0.11(0.04) & -- \\
MMD & 3.1(0.7) & 13(4) & 2.1(0.2) & 0.068(0.002) & 24(13) \\
MME & 4.4(0.7) & 5.5(0.7)& 6.8(0.7) & 0.15(0.04) & 80(23)  \\
IRS3-SW & 1.8(0.4) & 6(3) & 2.1(0.2) & 0.12(0.04) & 30(22) \\
\hline
\end{tabular}
\end{center}
\end{table*}

In Table~\ref{tab_coldensratio} we report the column densities of
\AMM , \nhdd\ and \H , and the column density ratios \nhdd -to-\AMM\ and \AMM -to-\H .
The \nhdd -to-\AMM\ ratio is of order 0.1 (from $\sim 0.07$ to 0.15), and does not change 
greatly from core to core. Such values are much larger than the
cosmic D/H ratio, estimated to be $\sim 10^{-5}$ (Linsky et al.~\citeyear{linsky}),
and comparable to those
measured towards low-mass pre--stellar cores (Roueff et al.~\citeyear{roueff}) 
and infrared dark clouds (Pillai et al.~\citeyear{pillai07}).
This implies that the deuteration 
in the cores of \ii\ is as high as in colder and more quiescent 
environments, despite the relatively higher gas temperature and turbulence,
and confirms previous findings in other dense cores associated with
intermediate- to high-mass star-forming regions (e.g.~Busquet et al.~\citeyear{busquet},
Pillai et al.~\citeyear{pillai11}).
Interestingly, relatively high values are found also close to the UC \HII\ region
associated with IRS~3, in MMB and IRS3-SW.
Because \nhdd\ is efficiently formed on dust grains, a strong UV radiation can
heat the dust and cause \nhdd\ evaporation thus increasing its abundance. On the other hand,
a strong UV radiation could (at least partly) decrease the abundance of \AMM , 
due to its interaction with H$^+$ to form NH$_3^+$ (e.g. Fuente et al.~\citeyear{fuente}).

Concerning the \AMM -to-\H\ ratio, we find the largest value in MME ($\sim 80$). 
The enhancement of the \AMM -to-\H\ ratio can be understood when
freeze-out of species heavier than He becomes important (see e.g.~Flower et al.~\citeyear{flower}),
so that it is expected to increase when the starless core gets closer to the 
onset of star formation. In this scenario, the fact that MME has the largest
\AMM -to-\H\ ratio suggests that this core could be close to the onset
of gravitational collapse, i.e. MME could be a candidate 
massive pre--stellar core.
However, Palau et al.~(\citeyear{palau07b}) measured with the SMA
a mass of only 0.7 \solm\ from the 1.3~mm continuum, while we find 23 \solm . 
This discrepancy likely comes from extended flux filtered out by the SMA,
which means that the core is quite flat and not centrally-peaked as
expected for a pre--stellar core. 

Based on the results of this work, we propose our final interpretation for the
nature of each one of the 3~mm condensations: 
\begin{itemize}
\item MMA is probably a protostar candidate. Although it does not show any
embedded infrared source, its relatively high \Tr , large line broadening, and 
\nhdd -to-\AMM\ lower than in other cores suggest
that this condensation is evolved. 
\item MME is likely a pre--stellar core, because it shows
high \nhdd -to-\AMM\ and \AMM -to-\H\ ratios, 
is more quiescent than MMA and it does not appear fragmented into smaller condensations 
when observed at higher angular resolution (Palau et al.~\citeyear{palau07b}).
Assuming a typical star formation efficiency of $\sim 30$\%, the core, the mass
of which is 23 \solm , has the potential to form an intermediate- to high-mass object.
\item the nature of MMB, MMC and MMD is less clear. Due to the
low \nhdd -to-\AMM\ and \AMM -to-\H\ ratios, MMD could be a protostellar
object, consistent with clear hints of contraction motions seen in the \AMM\ 
(1,1) spectrum, while for MMB we found hints of expansion due to 
asymmetric emission in the two inner satellites (see Sect.~\ref{res_linesvla}). 
Certainly, all condensations
are perturbed (MMB by the ionisation front from IRS~3, MMC by IRS~1 and the
outflow associated with it, MMD perhaps by a combination of both). Only
higher sensitivity and angular resolution observations will allow to better
understand the nature of these cores.
\end{itemize}

\subsection{Interaction of IRS~1 and IRS~3 with the dense gas: an expanding cavity}
\label{dis_cavity}

The most striking result of this work is the clear confirmation of a
cavity opened by IRS 3 in the molecular surrounding gas, and
a tight interaction between this cavity and the surrounding dense gas. 
We have found several evidences of this: (i) the
morphology of all the molecular tracers, especially in \cdh\ and \cthd ,
delineates a cavity around IRS~3 and the 1.3~cm continuum map resolving the ionised
gas perfectly matches the profile of the cavity;
(ii) the \AMM\ integrated intensity (2,2)/(1,1) ratios are large near IRS~3; 
(iii) the line widths are also large near IRS~3, specially in \cdh\ and \cthd ;
(iv) the position-velocity plot of \AMM\ shows a U-structure typical of an expanding
shell;
(v) in the MMB core we found hints of expansion in the \AMM\ (1,1) spectrum due to 
different intensity of the two inner satellites.
These evidences of such an interaction are shown for the first time in this work.

If we put together all the results obtained, we speculate about a possible
scenario that describes the star formation history in \ii : 
IRS~1 and IRS~3, both having bolometric luminosities of 
about 1000~\soll, seem to come from the same natal cloud while being clearly in 
different evolutionary stages, which points towards different generations of (intermediate-
to high-mass) star formation in \ii . In this context, IRS~3 could have induced the formation of IRS~1, 
as could be inducing star formation on the west (in MMA). On the other hand, 
in this bright-rimmed cloud the star formation probably has not been triggered by the UV radiation 
from the Cygnus OB2 association stars, because IRS~3, the massive star that
formed first, is relatively distant from the bright rim,
and the dense gas where we find the bulk of the current
star formation activity is around IRS~3 and away from the bright
rim. Therefore, the star formation seems to be dominated by IRS~3, 
which has been caught in the act of pushing away and disrupting its natal cloud.

\section{Summary and conclusions}
\label{conc}

The protocluster associated with the centre of \ii\ is an excellent
location where the interaction between evolved intermediate- and
high-mass protostellar objects
and dense (starless) cores can be studied.
We have derived the physical and chemical properties of the dense
gas by means of CARMA and VLA observations of the millimeter and centimeter
continuum, and of several molecular tracers (\cdh , \cthd , \AMM , \nhdd , \H ). 
Below, we summarise the main results.
\begin{itemize}
\item Morphologically, the dense gas is distributed in a filament oriented
east-west that passes in between IRS~1 and IRS~3, the two most massive and
evolved objects. 
We resolve the dense gas into five millimeter continuum condensations.
All of them show column densities consistent with potentially being the
birthplace of intermediate- to high-mass objects, although the masses
derived from continuum suggest that they can form intermediate-mass
objects more likely.  
\item We confirm the presence of an expanding cavity driven by IRS~3 
demonstrated mainly by the shape of the emission
in the two PDR tracers \cdh\ and \cthd , as well as by hints of expanding
motions from both the position-velocity diagrams and the
asymmetric intensity of the two inner satellites of the 
\AMM\ (1,1) line of the millimetre core closest to IRS~3 (MMB).
\item The non-thermal line widths across the filament indicate
that the gas kinematics is dominated by turbulence, similarly to other
intermediate- and high-mass star-forming regions and different from
low-mass dense starless cores.
\item The rotation and kinetic temperatures derived from ammonia are
on average larger than those typically found in cores associated
with low-mass star-forming regions, especially around the cavity
walls. The most massive and extended millimeter core, 
MME, shows physical and chemical signatures of an
intermediate- to high-mass pre--stellar core candidate.
\item We have better constrained the spectral index of the radio-continuum
emission associated with IRS~3, which turns out to be flat,
and thus the ionised gas comes from a region photoionised by 
the B2 ZAMS star at the centre of IRS~3.
\item the column density ratio \cdh /\cthd\ is of the
order of ~200-400 across the source and is higher where the dense gas 
is getting disrupted.
\item the deuterated fraction \nhdd -to-\AMM\ is of the order of 0.1
in all cores, as large as in low-mass pre--stellar cores and infrared
dark clouds. 
We find high levels of deuteration also close to the cavity driven by IRS~3.
We speculate that evaporation of \nhdd\ and \AMM\ destruction
caused by the UV radiation field can influence this high deuteration.
\end{itemize}

These findings undoubtedly confirm a tight interaction in \ii\ between the
most massive and evolved objects and the dense surrounding starless
cores in several respects (kinematics, temperature, chemical gradients),
and suggest that IRS~3 could have induced the formation of IRS~1 
and of the other gaseous condensations accumulated on the cavity
walls. However, the large-scale 
morphology of the molecular tracers suggests that we are likely seeing only a 
limited portion of a very extended gaseous filament. Only a large pc-scale 
mosaic will allow us to fully trace the distribution of the dense gas in 
the region and delineate a complete view of the core population.

\section*{Acknowledgments}

Support for CARMA construction was derived from the Gordon and Betty Moore
Foundation, the Kenneth T. and Eileen L. Norris Foundation, the James S. McDonnell
Foundation, the Associates of the California Institute of Technology, the University
of Chicago, the states of California, Illinois, and Maryland, and the National
Science Foundation. Ongoing CARMA development and operations are supported by the
National Science Foundation under a cooperative agreement, and by the CARMA partner
universities. We acknowledge support from the Owens Valley Radio Observatory, which
is supported by the National Science Foundation through grant AST 05-40399.
AP is grateful to Inma Sep\'ulveda for 
insightful discussions. AP is supported by the Spanish MICINN grant 
AYA2008-06189-C03 (co-funded with FEDER funds) and by a JAE-Doc CSIC 
fellowship co-funded with the European Social Fund
under the program `Junta para la Ampliaci\'on de Estudios'.
GB is funded by an Italian Space Agency (ASI) fellowship under contract 
number I/005/07/01. We are grateful to the anonymous referee for his/her valuable
comments and suggestions.


{}

\begin{appendix}

\renewcommand{\thefigure}{A-\arabic{figure}}
\renewcommand{\theequation}{A-\arabic{equation}}
\setcounter{figure}{0}  
\setcounter{equation}{0}  

\section*{Appendix A: Analysis for the case of
different $T_\mathrm{\lowercase{ex}}$ for the main and the
inner satellite lines of NH$_3$ $(1,1)$
}

\begin{figure}
\centering
\includegraphics[width=0.9\columnwidth]{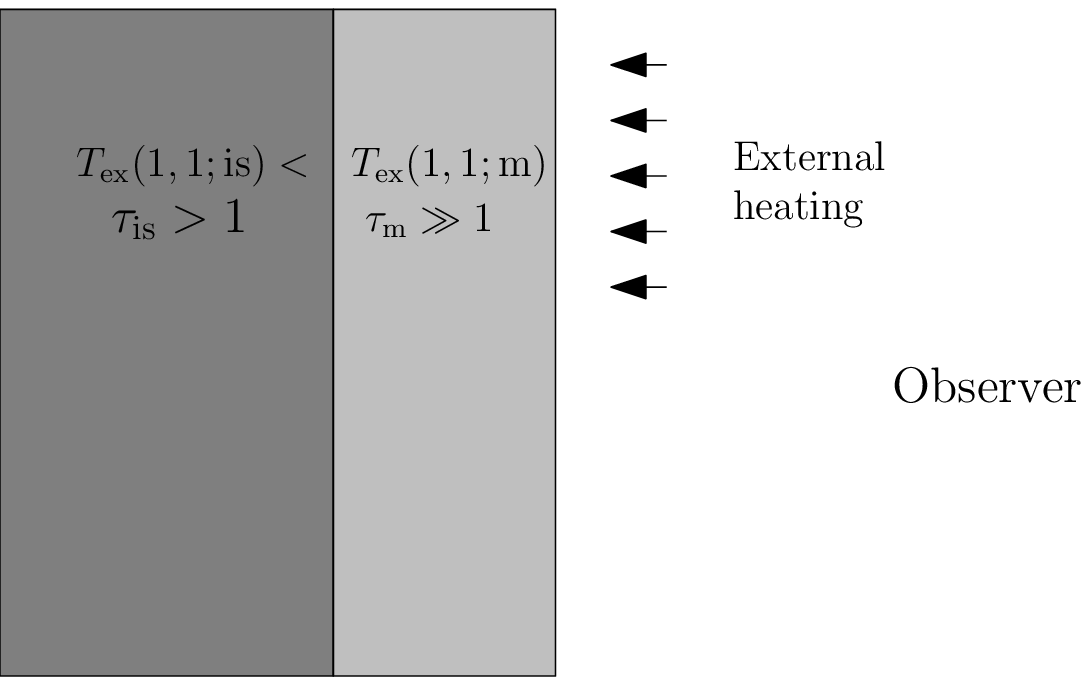}
\caption{Geometry of MMA}
\label{figappendix}
\end{figure}

Let us assume that the linewidth of the magnetic hyperfine components of the 
NH$_3$ $(1,1)$ inversion transition is large enough to make them unresolved, so
that only the five electric hyperfine lines, i.e.\  
one main line (``m''), 
two inner satellites (``is''),
and two outer satellites (``os''),
are resolved.
Assuming that the excitation temperature is $T_\mathrm{ex}\gg T_\mathrm{bg}$,
the ratio of intensities of the main line to the inner satellite line is
\begin{equation}
\frac{T_\mathrm{MB}(1,1;\mathrm{is})}{T_\mathrm{MB}(1,1;\mathrm{m})}=
\frac
{T_\mathrm{ex}(1,1;\mathrm{is})(1-e^{-\tau_\mathrm{is}})}
{T_\mathrm{ex}(1,1;\mathrm{m}) (1-e^{-\tau_\mathrm{m}})}.
\end{equation}
The usual assumption is that we are observing a homogeneous isothermal region,
so that both excitation temperatures, $T_\mathrm{ex}(1,1;\mathrm{is})$ and
$T_\mathrm{ex}(1,1;\mathrm{m})$, are equal, and that
$\tau_\mathrm{is}=0.28\,\tau_\mathrm{m}$. In this case,
\begin{equation}
0.28\leq
\frac{T_\mathrm{MB}(1,1;\mathrm{is})}{T_\mathrm{MB}(1,1;\mathrm{m})}
\leq1.
\end{equation}
The lower limit corresponds to the optically thin case, while the upper limit is
the optically thick case.

In Sect.\ 3.2.2 we have shown NH$_3$ $(1,1)$ spectra for each of the
3 mm continuum clumps. For the case of MMA,
the intensity ratio of the inner satellites and the main line is 
$0.23\pm0.05$, lower than the optically thin limit, $0.28$.

The assumption that the observed ratio is close to 0.28, and that the emission
in MMA is optically thin, leads to inconsistent results. For an optical depth of
the main line of $\tau_\mathrm{m}<0.1$, we obtain that $T_\mathrm{ex}>230$ K,
which is much higher than the kinetic temperature estimated from the
intensities ratio $T_\mathrm{MB}(2,2)/T_\mathrm{MB}(1,1)$ (see text),
$T_\mathrm{k}=22$ K. This result is improbable, since we expect 
the excitation temperature to be, in general, lower than the kinetic temperature.

The intensity of the $(1,1;\mathrm{m})$ is close to the kinetic temperature,
indicating that the optical depth of the main line is probably
$\tau_\mathrm{m}\gg1$. The optical depth of the satellite, however, can be
lower, so that both lines are tracing the emission of different layers of the
region observed: the main line, the outer layer facing the observer; and the
satellite line, a deeper layer of material (see Fig.\ \ref{figappendix}). 
The easiest explanation of the anomalous ratio 
$T_\mathrm{MB}(1,1;\mathrm{is})/T_\mathrm{MB}(1,1;\mathrm{m})$ is to assume that
the region is not isothermal, and that the two layers at different physical
depths, have different temperatures. So, the two excitation temperatures, 
$T_\mathrm{ex}(1,1;\mathrm{is})$ and $T_\mathrm{ex}(1,1;\mathrm{m})$, are not
equal. Thus, 
\begin{equation}
\frac{T_\mathrm{ex}(1,1;\mathrm{is})}{T_\mathrm{ex}(1,1;\mathrm{m})}
\frac{1-e^{-\tau_\mathrm{is}}}{1-e^{-\tau_\mathrm{m}}}=
\frac{T_\mathrm{MB}(1,1;\mathrm{is})}{T_\mathrm{MB}(1,1;\mathrm{m})}=
0.23. 
\end{equation}

Assuming that $T_\mathrm{ex}(1,1;\mathrm{m})$ is lower than $T_\mathrm{k}$,
\begin{equation}
T_\mathrm{MB}(1,1;\mathrm{m})\leq T_\mathrm{k} (1-e^{-\tau_\mathrm{m}}),
\end{equation}
giving that 
the optical depth of the main line must be $\tau_\mathrm{m}\ge3.1$. Thus,
taking into account that $\tau_\mathrm{is}=0.28\,\tau_\mathrm{m}$, we obtain
\begin{equation}
0.61\leq\frac{1-e^{-\tau_\mathrm{is}}}{1-e^{-\tau_\mathrm{m}}}\leq1,
\end{equation}
resulting in
\begin{equation}
0.23\leq
\frac{T_\mathrm{ex}(1,1;\mathrm{is})}{T_\mathrm{ex}(1,1;\mathrm{m})}
\leq0.38.
\end{equation}
The result is that the deeper layer traced by the satellite line is colder than
the outer layer traced by the main line. 
If we assume that the kinetic temperature of MMA, obtained from the ratio
$T_\mathrm{MB}(2,2)/T_\mathrm{MB}(1,1)$, is tracing the outer layer, the outer
layer temperature is 22 K, while the inner layer temperature is between 5 K
and 8 K.
The higher temperature of the outer layer is indicative of external heating, as
discussed in the text.

\end{appendix}

\label{lastpage}

\end{document}